\pdfoutput=1
\documentclass[final,5p,times,twocolumn]{elsarticle}

\usepackage{amsmath,amssymb}
\usepackage{graphicx}
\usepackage{booktabs}
\usepackage{multirow}
\usepackage{caption}
\usepackage{subcaption}
\usepackage{siunitx}
\usepackage{hyperref}
\usepackage{enumitem}
\usepackage{graphicx}
\usepackage{cleveref}
\usepackage{tabularx}
\graphicspath{{images/}}

\journal{Sustainable Energy, Grids and Networks}

\begin{document}
	
	\begin{frontmatter}
		
		\title{Channel Modeling of Single Wire Earth Return Networks for Narrowband Power Line Communication and Sensing: A Field-Validated High-Frequency Digital Twin}
		
		\author[1]{Wania Anoosh\corref{cor1}}
		\author[1]{Cagil Ozansoy}
		\author[1]{Douglas Gomes}
		\author[1]{Mike Faulkner}
		\author[1]{Kristi Beqirllari}
		
		\cortext[cor1]{Corresponding author: wania.anoosh@live.vu.edu.au}
		
		\affiliation[1]{organization={Institute for Sustainable Industries and Liveable Cities (ISILC), Victoria University},
			addressline={Footscray Park Campus},
			city={Melbourne},
			postcode={3011},
			country={Australia}}
		
		\begin{abstract}
			
			Upgrading Single-Wire Earth Return (SWER) networks for smart grid capabilities requires a reliable communications technol ogy. Narrowband Power Line Communication (NB-PLC) is a potential low cost solution. Real-world deployment is challenging due to the severe, frequency-dependent attenuation caused by complex earth-return paths and heterogeneous network infrastruc ture. To accurately characterize the communication channel, this paper develops a high-frequency (up to 300 kHz) digital twin of an operational SWER network. The digital twin integrates a segment-by-segment transmission line model with Vector Network Analyzer (VNA) measurements of physical grid hardware, replacing standard uniform assumptions with empirical component re sponses. Parametric sensitivity analysis demonstrates that distributed environmental factors, such as soil moisture and line sag, act as uniform magnitude offsets. Conversely, the conductor’s magnetic permeability and local injection-transformer impedances dictate the channel’s resonant spectral shape. Furthermore, cross-brand analysis proves that utilizing generic transformer models introduces significant prediction errors, confirming that accurate simulation requires manufacturer- and tap-specific data. Validated against in-situ field measurements from three transmitters, this digital twin replicates the path loss and dominant frequency-selective fading of the physical grid. Yielding a Root Mean Square Error (RMSE) between 4.65 dB and 9.73 dB across the three transmit paths, the model provides a practically reliable framework for deploying NB-PLC across rural SWER infrastructure.
			
		\end{abstract}
		
		
	\end{frontmatter}
	
	\section{Introduction}
	Modern smart grids rely on the continuous exchange of data to monitor and control the network. Utilities must upgrade older assets to include advanced metering, line monitoring, fault detection and control. Testing new communications and sensing waveforms directly on a live electrical network can be expensive,  involving significant down-time. Consequently, digital twins have become a necessary requirement for modern grid engineering. These virtual models allow researchers to safely simulate the behavior of communication signals under varying physical conditions before installing any actual hardware. While these digital twins are often standard for conventional multi-wire systems, a major gap remains when applying them to older, rural grid infrastructure.
	
	\indent Single-Wire Earth Return (SWER) distribution systems represent a unique class of power infrastructure that fundamentally differs from conventional multi-wire grids. Originally developed by Lloyd Mandeno in the 1920s to enable cost-effective rural electrification \cite{mandeno1947rural}, SWER utilizes a single medium-voltage conductor for power transmission, relying on the conductive earth itself to close the circuit and return load current to the source \cite{carson1926wave}. Recognized for their distinct simplicity and cost-effectiveness \cite{hosseinzadeh2011rural}, these systems remain the backbone of power distribution in vast rural areas, including Australia, New Zealand, and parts of Africa and South America. However, this minimalist design that ensures economic viability, also imposes an operational trade-off for contemporary smart grid integration. By eliminating the metallic neutral wire, SWER networks lack the stable signal return path typically used for grid telemetry and sensor data. Instead, communication signals are forced to travel through the earth itself. This presents a major challenge because the ground is an inconsistent medium; its impedance varies significantly depending on the local terrain, soil composition, and moisture levels \cite{mandeno1947rural}. Consequently, the global transition towards responsive smart grids has exposed a critical vulnerability: this legacy design effectively renders vast SWER networks as "dark assets", lacking the visibility and bidirectional communication required for modern fault localization and dynamic control. Given the sheer distance and cost of physically testing hardware on these remote lines, digital twins are the ideal approach, giving researchers a way to figure out exactly how communication signals will behave before any physical deployment.
	
	\begin{figure*}[t]
		\centering
		\includegraphics[width=0.95\textwidth]{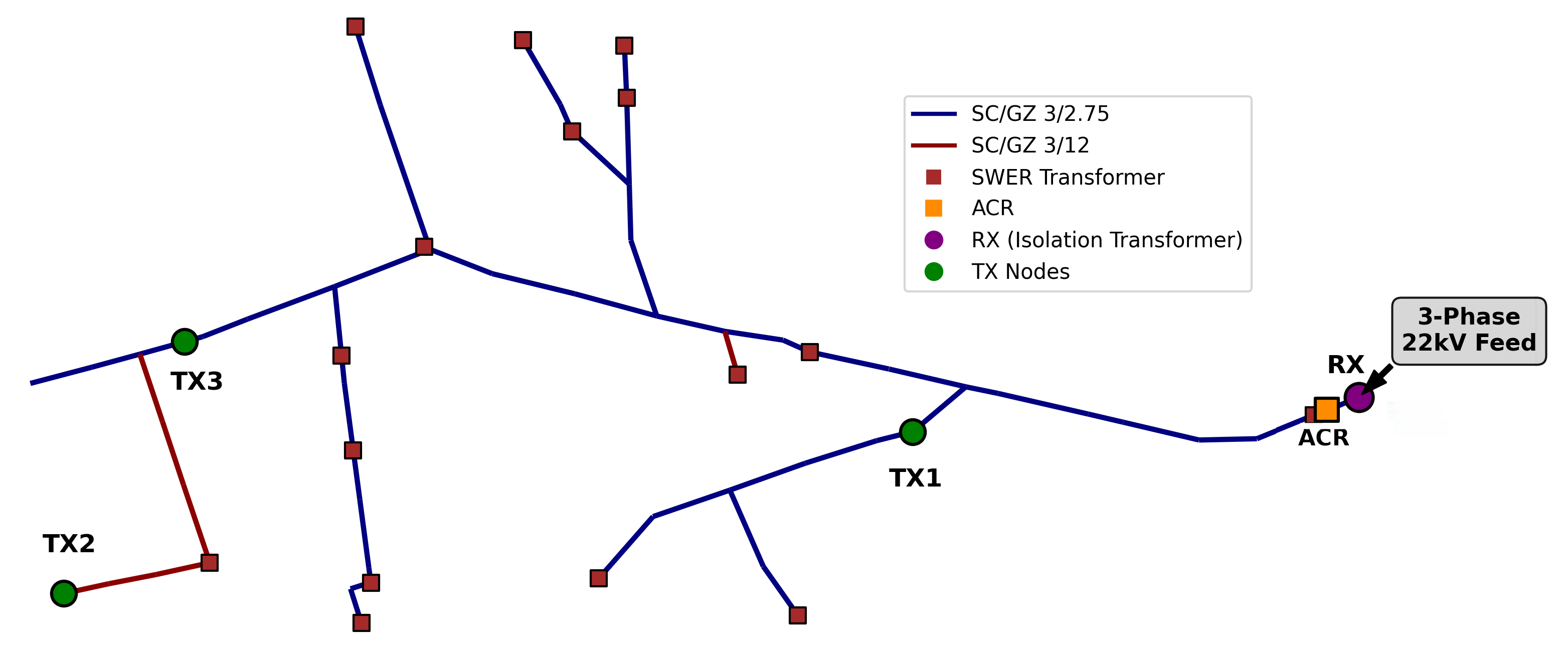}
		\caption{Network schematic diagram of the 15.31 km SWER network in Victoria, Australia, illustrating the 3-phase isolation feed and the distributed MV-to-LV transformers at individual customer premises.}
		\label{fig:SpringhillKML}
	\end{figure*}	
	
	\indent Narrowband Power Line Communication (NB-PLC) offers a practical way to modernize this aging infrastructure by using existing power lines for data transmission, avoiding the high cost of deploying new fiber or wireless networks \cite{galli2011grid}. In the context of SWER networks, which often span tens of kilometers across remote terrain, NB-PLC is far more suitable than Broadband PLC (BB-PLC). While BB-PLC provides higher data rates, its signal attenuates rapidly over long distances, which would necessitate an impractical number of repeaters to cover typical SWER transmission distances \cite{ancillotti2013role}. NB-PLC, on the other hand, operates at lower frequencies with much better propagation. This stability allows it to support critical applications, ranging from basic smart metering (AMI) \cite{gungor2011smart} to advanced grid monitoring like remote fault detection and demand-side management \cite{fernandez2025leveraging}.
	
	\indent Despite this potential, the deployment of PLC over geographically extensive SWER infrastructure is inherently challenging, resulting in performance that is unpredictable and highly site-dependent \cite{tonello2015considerations}. Standard power system analysis treats SWER conductors as constant parameters at fundamental frequency (50/60 Hz). However, effective communication relies on the NB-PLC spectrum (9 kHz–500 kHz), where these static assumptions no longer apply. Operating in this higher frequency band necessitates treating the conductors as complex transmission lines characterized by frequency-dependent parameters \cite{malik2018inclusion}. This modeling requirement is dictated by the unique physical constraints of SWER. 
	
	\indent Balanced three-phase systems benefit from low-loss inter-conductor transmission modes \cite{taylor1995digital}, while SWER propagation is inherently constrained to a single conductor and a lossy earth-return path. Consequently, these line parameters are driven by two main factors: the skin effect and permeability within the metal wire, and the fluctuating conductivity and permittivity of the surrounding soil \cite{carson1926wave,papadopoulos2020closed}. Seminal studies by Kikkert and Nkom have significantly advanced the field by developing rigorous frequency-dependent models to characterize SWER transmission line segments and coupling components \cite{kikkert2009radiation, nkom2017narrowband}. Nkom expanded on this by integrating these component models to successfully simulate complete communication channels \cite{nkom2016unified}. However, these efforts largely assumed a consistent, idealized infrastructure. They did not establish a framework for simulating signal propagation across a real-world, multi-branched network built from heterogeneous and non-standardized components. 
	
	\indent While detailed models for individual components already exist, they are almost always applied to perfect and uniform network layouts. There is a critical knowledge gap here because no published work has systematically evaluated how a realistic mixture of non-standardized components would change the signal propagation across a whole network. This paper aims to fill this gap by developing a segment-by-segment co-simulation framework capable of characterizing the SWER infrastructure as a high-frequency communication channel. A further strength herein is that the proposed model is validated against end-to-end field measurements, accurately capturing the cumulative impact of network variability on signal attenuation. This is achieved by integrating:		
	\begin{enumerate}[noitemsep]
		\item A \textbf{segment-by-segment} transmission line model, based on the fundamental work of Nkom and Sunde \cite{nkom2017narrowband, 1970304959890429482, rachidi2008electromagnetic}, which calculates frequency-dependent per-unit-length (p.u.l.) parameters for different conductor types. 
		
		\item \textbf{Vector Network Analyzer (VNA)} measured Scattering Parameters (S-parameters) that characterize the high-frequency response of real-world discrete components, including distribution transformers, line filters, coupling capacitors, and Automatic Circuit Reclosers (ACRs).
	\end{enumerate}
	
	\begin{figure*}[t]
		\centering
		\begin{subfigure}[b]{0.49\textwidth}
			\includegraphics[width=\textwidth]{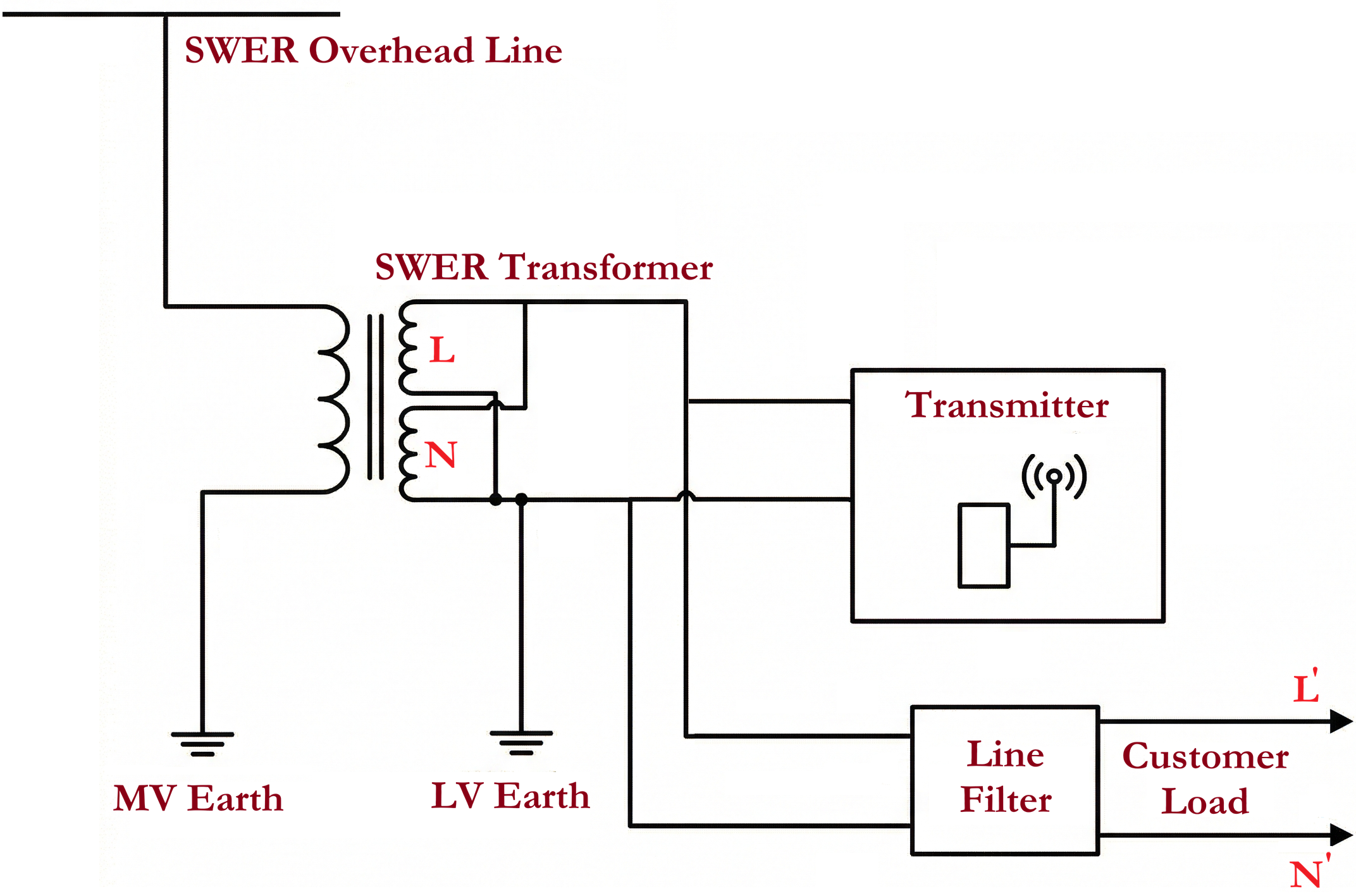} 
			\caption{LV Signal Injection Side (Transmitter)}
		\end{subfigure}
		\hfill
		\begin{subfigure}[b]{0.49\textwidth}
			\includegraphics[width=\textwidth]{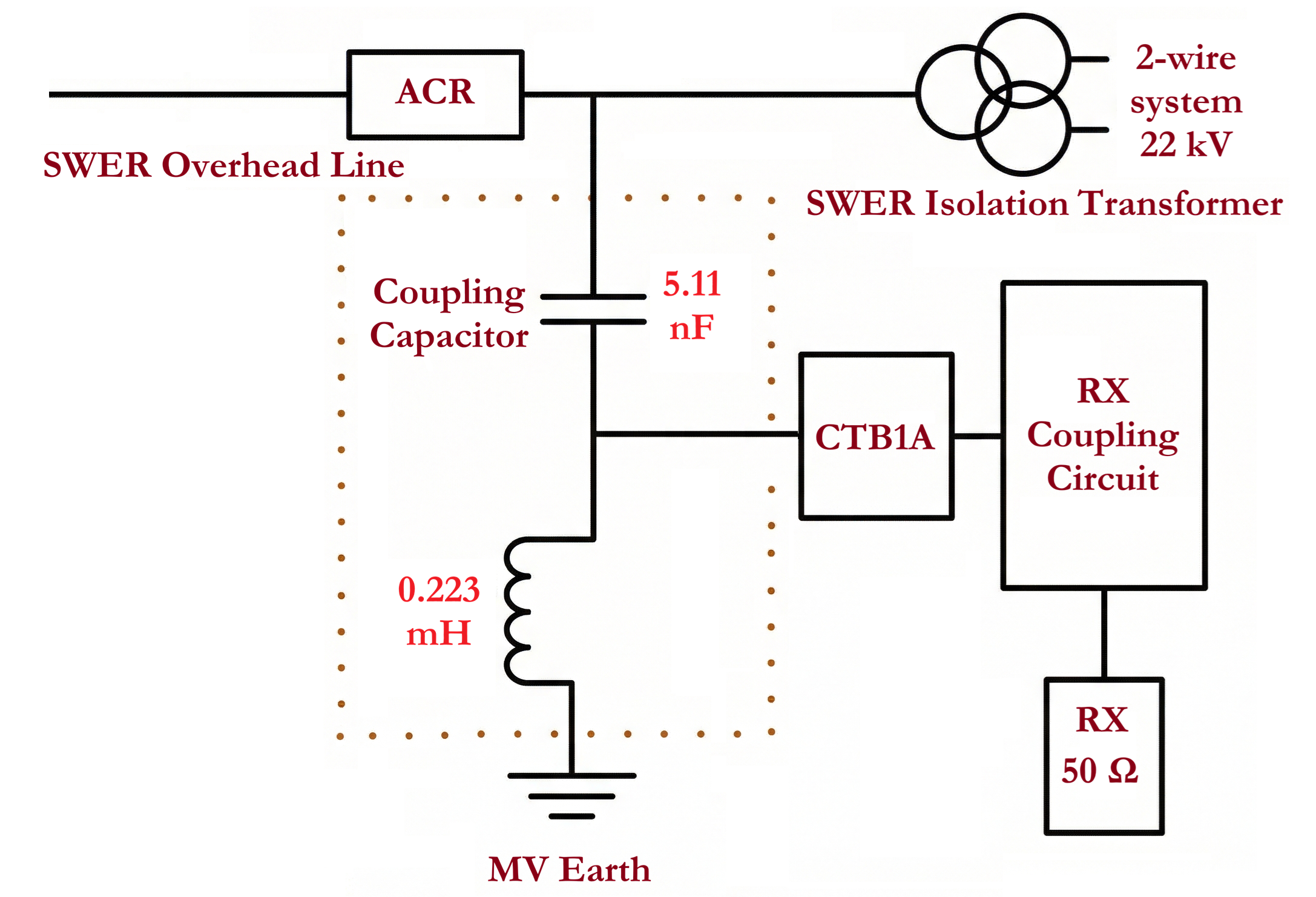}  
			\caption{MV Signal Extraction Side (Receiver)}
		\end{subfigure}
		\caption{Circuit topologies of transmitter and receiver side of the SWER channel. The transmitter configuration (a) utilizes a shunt connection with an LV line
			filter to isolate the customer load, while the receiver configuration (b) employs a series coupling capacitor and an ACR for medium-voltage signal extraction and
			network isolation.}
		\label{fig:TX_vs_RX_Side}
	\end{figure*}
	
	\indent To demonstrate the practical accuracy of this framework, a simulation twin was developed for an operational rural SWER network located in Victoria, Australia. Figure \ref{fig:SpringhillKML} provides the geospatial layout of the physical network, KML-generated plot that details the transmission line paths, the placement of the SWER transformers, and the specific locations of the communication nodes. This system comprises three transmitter units installed on the low-voltage (LV) side of distribution transformers, marked as TX1, TX2, and TX3 respectively. Notably, at the TX1 and TX3 sites, the transmitters are deployed in parallel with LV line filters to isolate the communication signals from downstream customer loads (Figure \ref{fig:TX_vs_RX_Side}a), whereas the TX2 site operates without a filter. These units inject unique, identifiable signals which are monitored by a single receiver (RX) co-located with an ACR at the network's isolation point. By coupling the receiver to the high-voltage line via a coupling capacitor, this setup captures the full end-to-end channel response (Figure \ref{fig:TX_vs_RX_Side}b). The rigorous development and validation of this Digital Twin against the in-situ field measurements constitutes the central contribution of this work.
	
	\indent Beyond validating the simulation framework, this study uses parametric sensitivity analysis to isolate the physical mechanisms behind SWER channel behavior. The results show that the injection transformer and its tap setting are the dominant factors controlling signal magnitude. When looking at the spectral shape, the frequency response is highly sensitive to the magnetic permeability ($\mu_r$) of the conductor, which acts like a tunable bandwidth filter. Finally, this analysis also demonstrates that network heterogeneity is a critical variable. Specific component designs from different manufacturers introduce unique attenuation patterns that simplified models fail to predict. By resolving these factors, this research provides a physically grounded diagnostic of signal propagation in complex SWER grids.		
	
	\indent The remainder of this paper is organized as follows. Section 2 reviews the existing literature on SWER channel modeling and the application of sensitivity analysis in power systems. Section 3 details the methodology used to build the co-simulation framework, including the mathematical line models and component measurements. Section 4 discusses the results of parametric sensitivity analysis, interprets the physical drivers of channel	attenuation, and compares the spectral response of the developed simulation twin against the field measurements from the SWER network. Finally, Section 5 summarizes the key findings and concludes the paper.
	
	\section{Related Work}
	The modeling of SWER networks as high-frequency PLC channels sits at the intersection of classical power system analysis and microwave network theory. The literature remains fragmented and can be broadly categorized into three distinct areas: (1) frequency-dependent transmission line modeling, (2) high-frequency characterization of discrete components, and (3) parametric sensitivity analysis in power networks.
	
	\subsection{Frequency-Dependent Line Modeling}
	Accurately modeling the SWER PLC channel requires explicitly calculating the earth return impedance, as its value changes across the communication band. The theoretical foundation for this was established by Carson \cite{carson1926wave} and was later refined by Sunde \cite{1970304959890429482, rachidi2008electromagnetic}. These classical works demonstrated that the p.u.l. parameters of the line, particularly its series impedance, are not constant. The ground path acts as a complex, frequency-dependent impedance, governed by the soil's resistivity ($\rho$) and permittivity ($\epsilon_r$). While Carson’s work assumed uniform ground conductivity, Sunde expanded this to wider frequency ranges by accounting for displacement currents in the soil, a critical factor for PLC applications.		
	
	\indent In the context of modern smart grids, Nkom et al. adapted these classical formulations specifically for narrowband PLC on SWER, providing rigorous methods to calculate p.u.l. parameters for Aluminium Conductor Steel Reinforced (ACSR) conductors \cite{nkom2017narrowband}. Nkom extended this work to characterize the distribution transformer as a dynamic PLC channel, identifying that energization levels and dielectric polarization significantly impact signal integrity \cite{nkom2017impact}.		
	
	\indent However, while Nkom \cite{nkom2016unified} successfully integrated line models with discrete transformer models to analyze attenuation across fundamental network links, this body of work remains focused on fundamental, point-to-point topologies. This paper builds directly on Nkom's foundational work by expanding the scope to a real-world SWER grid, where signal energy splits and reflects across a complex topology of branches, spurs, and heterogeneous infrastructure. Clear analytical boundaries were necessary to simulate a network on the scale of Figure \ref{fig:SpringhillKML}. To keep the model manageable, downstream MV-to-LV transformers were standardized to a fixed tap setting, and uniform material and ground constants were applied globally.
	
	\subsection{High-Frequency Component Characterization}
	Discrete network infrastructure, particularly distribution transformers, introduces complex impedance discontinuities that are negligible at power frequencies but critical for communications. In the PLC frequency band (9–500 kHz), transformers behave not as simple inductive elements but as complex, multi-resonant circuits dominated by leakage inductance and inter-winding capacitance \cite{gustavsen2004wide}. To capture this behavior, Kikkert treated these components as multi-port microwave networks, utilizing VNA measurements to derive S-parameter models \cite{kikkert2010modelling}. While these isolated models are highly accurate, they do not account for the cumulative interactions that occur when multiple components are cascaded across a physical distribution grid.
	
	\indent When modeling large-scale networks, researchers frequently approximate auxiliary components and transformers as generic functional blocks. Applying a standardized 25 kVA transformer model is often a practical necessity to keep complex network simulations computationally viable. However, relying solely on this generalization can mask critical high-frequency variations, particularly at the signal injection point. Foundational studies by Zimmermann and Dostert \cite{zimmermann2002multipath} successfully characterize powerline networks using measurement-based parametric models. These top-down statistical methods are excellent for mapping the overall end-to-end multipath environment. Yet, because they mathematically aggregate the channel into a general propagation medium, they inherently abstract the specific, frequency-dependent impacts of discrete hardware like line filters, coupling capacitors and ACRs. 
	
	\indent To complement these top-down models, the literature requires a bottom-up framework that dynamically integrates lab-measured component parameters into a full-scale network topology, accurately linking individual physical assets to the complete system's behavior. In this study, discrete equipment including distribution transformers, ACRs, and line filters were treated as multi-port microwave networks and measured using a VNA across a 1 to 1601 kHz sweep. All measurements were conducted on de-energized equipment in a controlled laboratory setting. 
	
	\subsection{Parametric Sensitivity Analysis in Power Systems}
	To resolve the persistent discrepancies between theoretical models and field behavior, recent research in adjacent power domains has adopted parametric sensitivity analysis as a primary diagnostic tool \cite{mitchell2011modeling}. This methodology moves beyond simple error minimization, instead quantifying how specific uncertainties in physical parameters propagate to drive system-level deviations. By establishing a deterministic mapping between physical parameter variations and spectral shifts, this approach allows for the isolation of critical variables that define the high-frequency response. In the specific context of 50 Hz SWER systems, Bakkabulindi \cite{bakkabulindi2012planning} used sensitivity analysis to demonstrate that the planning and operational validity of the network is heavily dependent on the inherent variability of the earth return path. 
	
	\indent While sensitivity analysis typically focuses on soil parameters, the specific material composition of SWER conductors introduces a critical source of uncertainty. A defining characteristic of these networks is the dominant use of Steel Core Galvanized Zinc (SC/GZ) conductors, distinct from the ACSR lines used in standard grids. Since SC/GZ is ferromagnetic, its relative permeability ($\mu_r$) is significantly higher than that of aluminum and behaves non-linearly. Morgan \cite{morgan1965electrical} established that permeability fluctuates with line current, while Meyberg  \cite{meyberg2021magnetic} identified further variations due to temperature, composition and manufacturing process. This inherent variability of $\mu_r$ necessitates a parametric sensitivity analysis to quantify exactly how this magnetic uncertainty impacts channel performance. This uncertainty further extends to distribution transformers, where off-nominal tap settings introduce variable impedance discontinuities that significantly shift the attenuation profile. In contrast to studies that treat these parameters as fixed constants, this research applies parametric sensitivity analysis to diagnose the unique physics of the SWER PLC channel.
	
	\indent Ultimately, the existing literature leaves three major gaps that limit the accurate characterization of the communication channel across real-world SWER grids. First, current transmission line models focus on simple, point-to-point links rather than the multi-branched reality of mixed conductors and equipment. Second, while individual devices have been measured via VNA, these isolated profiles have not been integrated into a full-scale simulation to show how cascaded components interact. Finally, while sensitivity analysis is used for variables like soil resistivity, previous work has not clearly separated the physical factors that simply lower the signal's magnitude from those that completely alter its spectral shape.
	
	\section{Modeling Framework}
	A multi-domain co-simulation framework was developed to construct the digital twin of a physical SWER network. This work develops a novel segment-by-segment, frequency-dependent model of a complete SWER network that explicitly incorporates conductor composition, earth return characteristics, and physical geometry. The framework integrates analytical formulations (MATLAB) for transmission line parameters with distributed-parameter circuit simulations (Microwave AWR), incorporating VNA-measured S-parameters for all discrete network components. 
	
	\subsection{Frequency-Dependent Transmission Line Model}
	The SWER transmission line is the dominant distributed element of the channel. Its electrical parameters exhibit significant frequency dependence, due to the skin effect within the ferromagnetic SC/GZ conductors and the complex impedance of the earth return path. To capture these behaviours, this study uses the fundamental formulations derived by Nkom and Sunde \cite{nkom2017narrowband, 1970304959890429482}. The p.u.l. effective resistance, $R$, incorporates the skin depth ($\delta = \sqrt{\rho / \pi \mu f}$) \cite{kikkert2011calculating} and the conductor's multi-strand geometry:       		
	
	\begin{equation}
		R = \sqrt{\frac{R_{\text{Cdc}}^{2} \cdot r_{s}^{2}}{4 \cdot \delta^{2}} + R_{\text{Wdc}}^{2}} \quad \left( \frac{\Omega}{\text{m}} \right)
	\end{equation}
	
	where $R_{\text{Cdc}}$ is the DC resistance of a solid, homogeneous conductor of the same base material (e.g., aluminum or steel) \cite{kikkert2011effect}, $R_{\text{Wdc}}$ is the actual DC resistance of the conductor which differs from $R_{\text{Cdc}}$ in a non-homogeneous conductor, and $r_s$ is the radius of each strand. The total p.u.l. resistance $R_{\text{total}}$ also incorporates frequency-dependent p.u.l. radiation resistance $R_{\text{r}}$ \cite{kikkert2011effect}.
	
	\indent The total inductance, L, is modeled as the sum of the external inductance ($L_{ext}$), governed by the line geometry, and the internal inductance ($L_{int}$), which is highly sensitive to the relative permeability of the ferromagnetic steel core:
	
	\begin{equation}
		L_{\text{ext}} = \frac{\mu_{0} \cdot \mu_{R\text{-air}}}{2 \pi} \ln\left(\frac{D}{r}\right) \quad \left( \frac{\text{H}}{\text{m}} \right)
	\end{equation}
	
	\begin{equation}
		L_{\text{int}} = \frac{\mu_{0} \cdot \mu_{R\text{-steel}}}{\sqrt{\frac{r^{2}}{4 \delta^{2}} + 1}} \quad \left( \frac{\text{H}}{\text{m}} \right)
	\end{equation}
	
	where $\mu_{\text{o}}$ is the permeability constant, $\mu_{R-\text{air}}$ and $\mu_{R-\text{steel}}$ is the relative permeability of air and steel respectively, and $r$ is the conductor's radius. Following the Method of Images (MOI) as applied in the SWER PLC modeling framework by Nkom \cite{nkom2017narrowband}, D is the distance between the conductor and its theoretical image below the ground plane, such that $D = 2h$, where h is the height of the overhead conductor. 
	
	\indent The per-meter capacitance between SWER line and earth is calculated using the following equation:
	\begin{equation}
		C = \frac{2 \pi \cdot \epsilon_0 \cdot \epsilon_{r(\text{A})} }{\ln\left(\frac{D}{r}\right)} \quad \left( \frac{\text{F}}{\text{m}} \right)
	\end{equation}
	
	where $\epsilon_{\text{0}}$ is the permittivity of free space and $\epsilon_{r(A)}$ is the relative permittivity of air.
	
	\begin{figure}[t]
		\centering
		\includegraphics[width=\columnwidth]{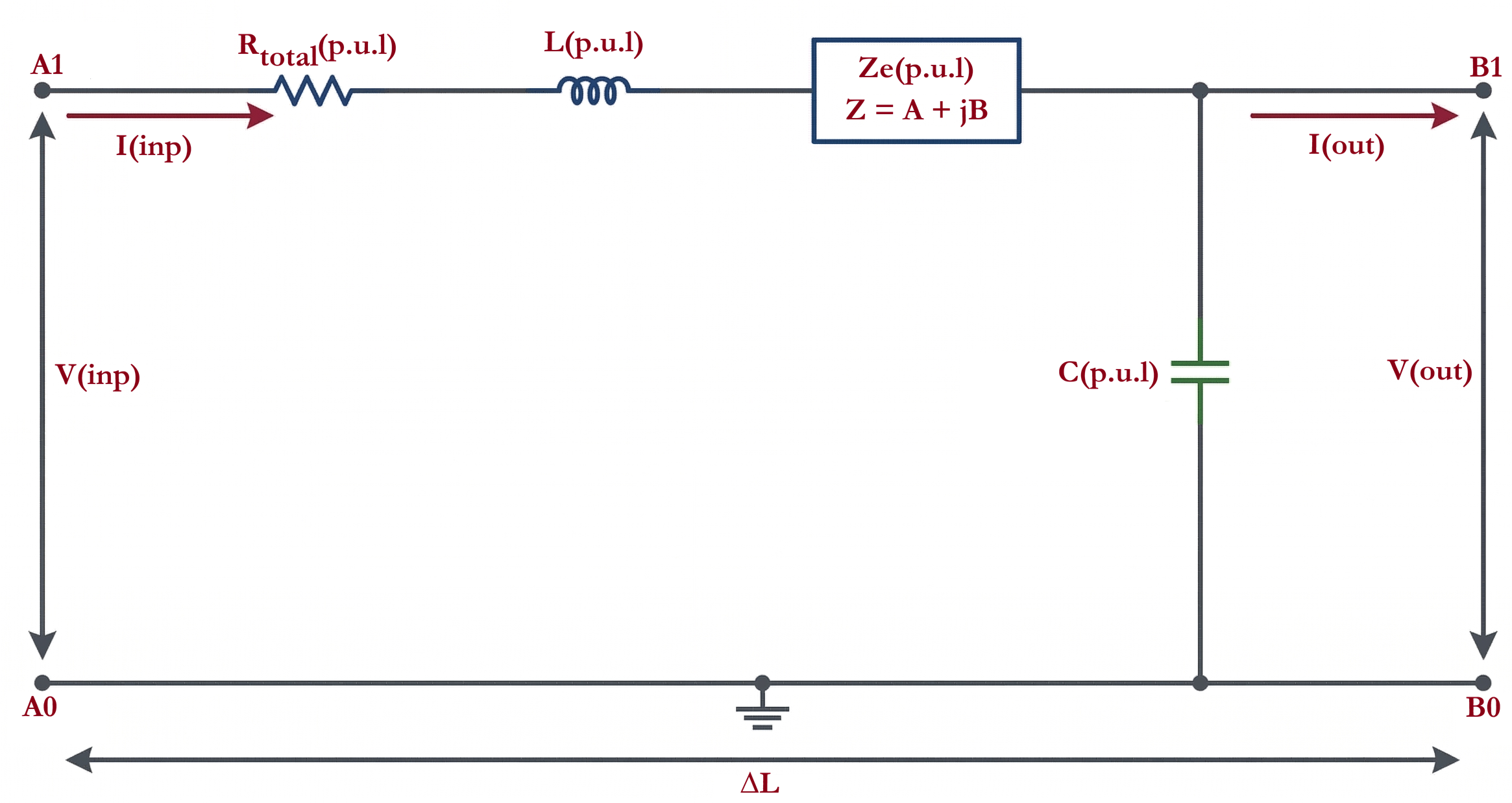}
		\caption{Analytical per-unit-length (p.u.l.) circuit model of the SWER transmission line segment ($\Delta L$)}
		\label{fig:line_model_circuit}
	\end{figure}
	
	\begin{figure}[t]
		\centering
		\includegraphics[width=\columnwidth]{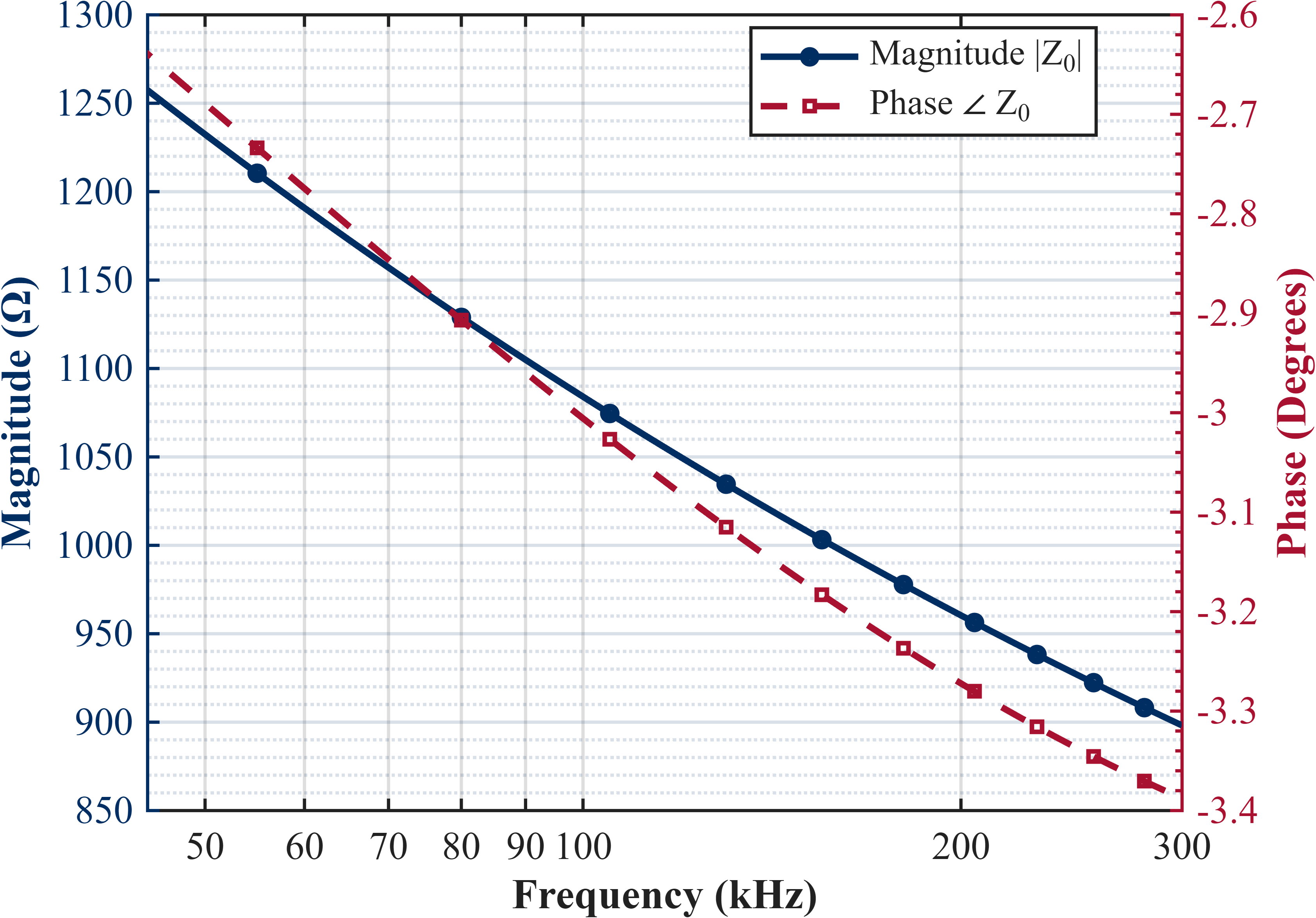}
		\caption{Magnitude and phase of the characteristic impedance ($Z_0$) across the operating frequency band for the baseline SWER configuration.}
		\label{fig:Z0_plot}
	\end{figure}
	
	\indent Crucially, the earth return impedance $Z_e$ is modeled using Sunde’s logarithmic approximation, which remains valid across the PLC band by accounting for displacement currents and the complex permittivity of the soil:
	\begin{equation}
		Z_{\text{e}} = j f \mu_0 \ln \left[ \frac{1 + h \gamma_{\text{(e)}}}{h \gamma_{\text{(e)}}} \right] \quad \left( \frac{\Omega}{\text{m}} \right)
	\end{equation}
	
	where f is the frequency in Hz, $\gamma_{\text{e}}$ is the earth propagation constant derived from soil resistivity ($\rho_e$) and relative permittivity ($\epsilon_r$), and is given by:
	\begin{equation}
		\gamma_{\text{(e)}} = \sqrt{j 2 \pi f \cdot \mu_0 \left( \left[ \frac{1}{\rho_{\text{e}}} \right] + j 2 \pi f \cdot \epsilon_0 \cdot \epsilon_{\text{r(e)}} \right)} \quad \left( m^{-1} \right)
	\end{equation} 
	
	\indent Figure \ref{fig:line_model_circuit} illustrates how these individual components are combined into an analytical p.u.l. circuit model for the Microwave AWR digital twin. As shown, the frequency-dependent conductor properties and the earth return impedance are integrated into the series path, while the capacitive coupling forms the shunt path to ground. The total complex series impedance ($Z_{\text{series}}$) and total shunt admittance ($Y_{\text{shunt}}$) of the SWER line are defined as:
	
	\begin{equation}
		Z_{\text{series}} = R_{\text{total}} + j 2 \pi f (L_{\text{ext}} + L_{\text{int}}) + Z_{\text{e}} \quad \left( \frac{\Omega}{\text{m}} \right)
	\end{equation}
	
	\begin{equation}
		Y_{\text{shunt}} = G + j 2 \pi f C \quad \left( \frac{\text{S}}{\text{m}} \right)
	\end{equation}
	
	where the shunt conductance, $G$, is assumed to be negligible for overhead conductors. From these distributed parameters, the characteristic impedance ($Z_0$) and the propagation constant ($\gamma_{\text{line}}$) of the SWER channel are calculated:
	
	\begin{equation}
		Z_0 = \sqrt{\frac{Z_{\text{series}}}{Y_{\text{shunt}}}} \quad (\Omega)
	\end{equation}
	
	\begin{equation}
		\gamma_{\text{line}} = \sqrt{Z_{\text{series}} \cdot Y_{\text{shunt}}} = \alpha + j\beta \quad \left( m^{-1} \right)
	\end{equation}		
	
	\begin{figure*}[t]
		\centering
		\begin{subfigure}[b]{0.49\textwidth}
			\includegraphics[width=\textwidth]{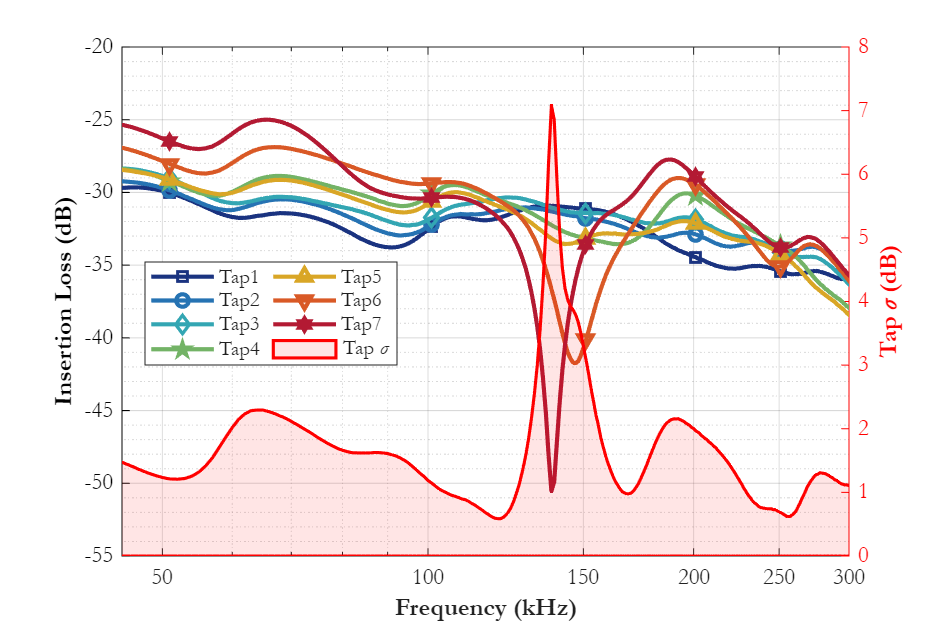} 
			\caption{Manufacturer A}
		\end{subfigure}
		\hfill
		\begin{subfigure}[b]{0.49\textwidth}
			\includegraphics[width=\textwidth]{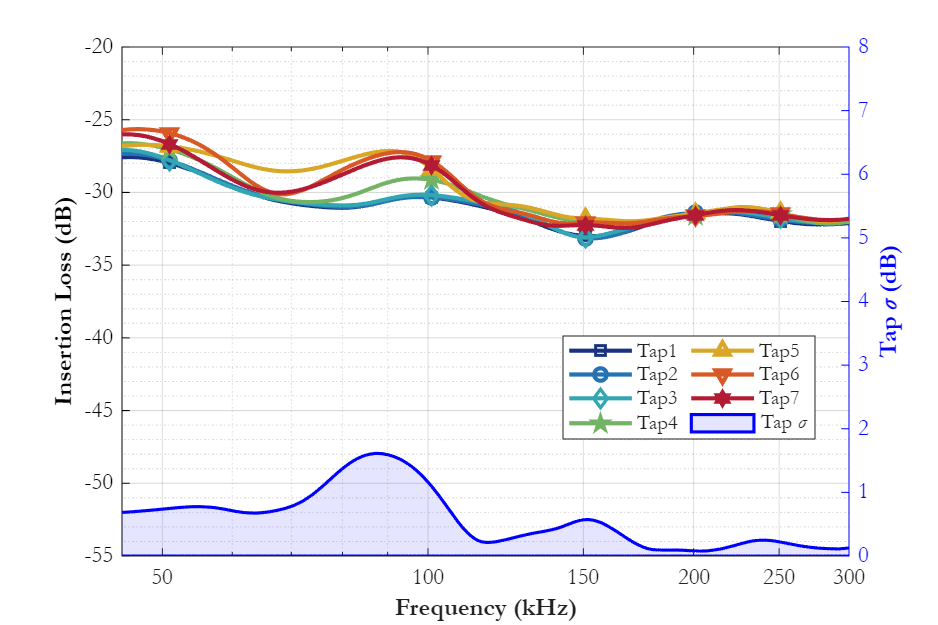}  
			\caption{Manufacturer B}
		\end{subfigure}
		\caption{Measured Insertion Loss ($S_{21}$) for 25 kVA SWER transformers from two different manufacturers across operational Taps 1--7.}
		\label{fig:S21_Comparison}
	\end{figure*}
	
	\indent Figure \ref{fig:Z0_plot} illustrates the magnitude and phase of $Z_0$ across the operational frequency band. The high-frequency attenuation constant ($\alpha$) is fundamentally governed by the relationship between the total series resistance and the characteristic impedance:
	
	\begin{equation}
		\alpha \approx \frac{R_{\text{total}}}{2 |Z_0|} \quad \left( \frac{\text{Np}}{\text{m}} \right)
	\end{equation}
	
	\indent This relationship proves that path loss is not dictated by series resistance alone, but scales inversely with characteristic impedance. Consequently, physical variations that alter distributed inductance or capacitance will shift $Z_0$ and fundamentally reshape the attenuation profile. This mathematical relationship forms the basis for the parametric sensitivity analysis presented in Section 4.
	
	\subsection{Component Characterization via VNA Measurements}
	In a real-world SWER network, discrete components such as distribution transformers, ACRs, and line filters act as significant impedance discontinuities. To accurately capture their behavior, these components were treated as multi-port microwave networks. S-parameters were measured for each device using a VNA. These empirical blocks replace idealized circuit elements, embedding real parasitic behavior and inter-winding coupling effects.	
	
	\begin{figure}[t]
		\centering			
		\includegraphics[width=\columnwidth]{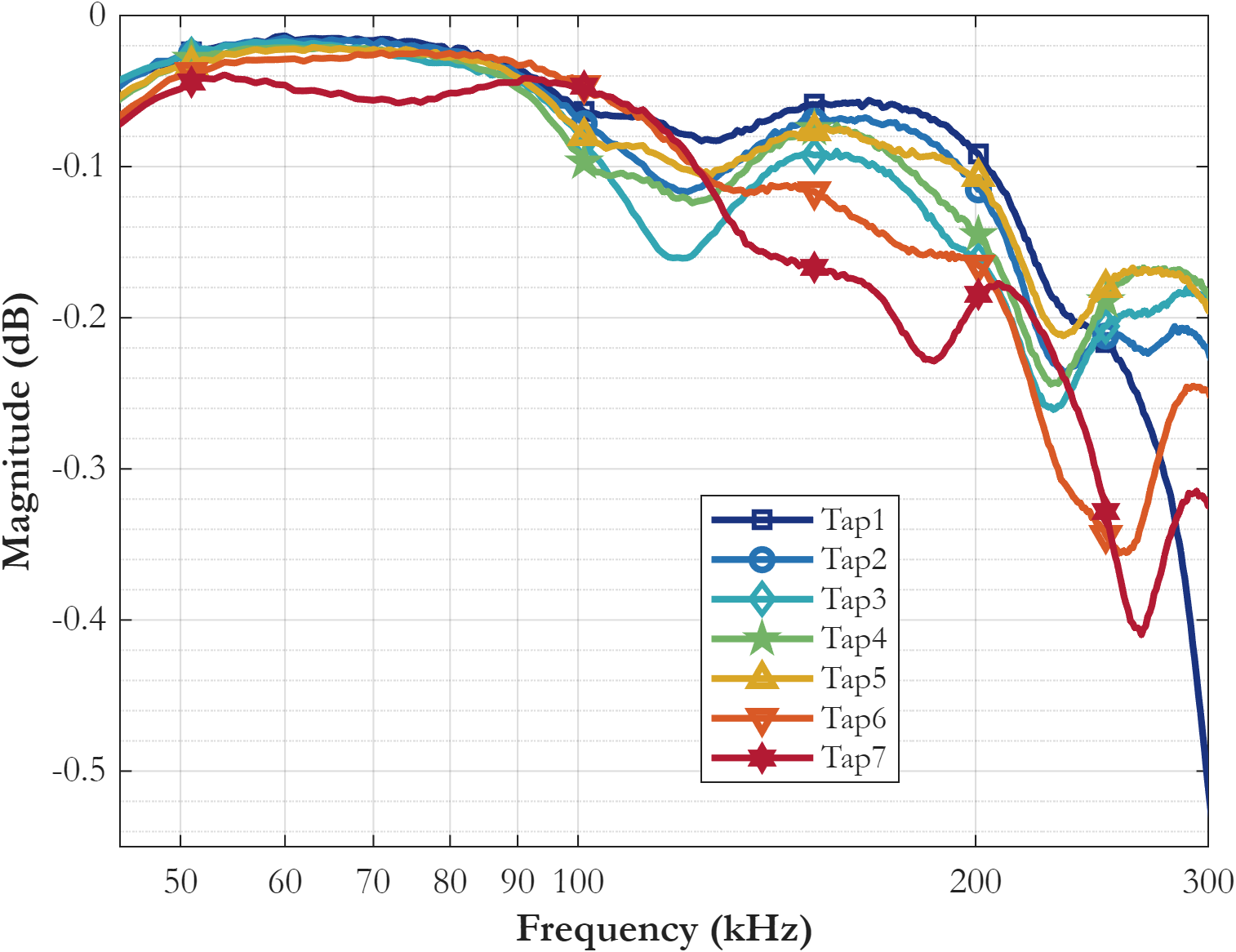} 
		\caption{Measured reflection ($S_{22}$) of 25 kVA SWER transformer from Manufacturer A across operational taps 1–7.}
		\label{fig:S22_Comparison}
	\end{figure}
	
	\subsubsection{Distribution Transformer Heterogeneity}		
	The SWER transformers act as complex resonant structures; their internal winding geometry impacts where these resonances occur. They constitute one of the primary source of signal attenuation and reflection within the PLC band. Distribution transformers terminate the MV spur lines and are responsible for the reflections occurring at the ends of the network branches. To quantify the impact of equipment diversity, insertion loss ($S_{21}$) measurements (Figure \ref{fig:S21_Comparison}) were conducted on 25 kVA transformers from two different manufacturers (designated Manufacturer A and Manufacturer B) across all seven operational tap settings. Additionally, the return loss ($S_{22}$) was measured for Manufacturer A (Figure \ref{fig:S22_Comparison}) to determine the proportion of the signal reflected or absorbed by the impedance mismatch on the MV side.
	
	\indent Looking at the $S_{22}$ profile for Manufacturer A (Figure \ref{fig:S22_Comparison}), the response remains remarkably flat and close to 0 dB across the band, indicating that almost the entire signal is reflected due to the transformer's high input impedance compared to 50 $\Omega$ VNA ports. It is important to note that the actual impedance a transformer "sees" depends on its location in the network. If it is positioned at the end of a line, it sees the full characteristic impedance. If it is located in the middle of a line or near a T-junction, it sees a lower impedance. An $S_{22}$ value below 0.15 dB indicates that over 70\% of the signal is reflected by terminating transformers, with even higher reflection rates in non-terminating setups. 
	
	\begin{table}[t]
		\centering
		\caption{Statistical Deviation in Transformer’s $S_{21}$ Parameters (45--300 kHz)}
		\label{tab:freq_stats}
		\small
		\begin{tabularx}{\columnwidth}{@{}Xccc@{}}
			\toprule[1.5pt]
			\textbf{Metric (dB)} & \textbf{Mfr A} & \textbf{Mfr B} & \textbf{A+B} \\ 
			\midrule[1.5pt]
			Mean  Attenuation @ 45kHz & -28.1 & -26.7 & -27.4 \\
			Slope (dB/dec) & -7.7 & -5.0 & -6.3 \\
			Mean $\sigma$ & 1.6 & 0.4 & 1.5 \\
			$\sigma$ @ 50kHz & 1.2 & 0.7 & 1.3 \\
			$\sigma$ @ 150kHz & 3.3 & 0.6 & 2.4 \\
			$\sigma$ @ 250kHz & 0.7 & 0.2 & 1.5 \\
			Max Variation & 19.6 & 3.5 & 19.6 \\ 
			\bottomrule[1.5pt]
		\end{tabularx}
	\end{table}
	
	\indent Table \ref{tab:freq_stats} summarises the variation between the taps, within and across both transformers. Manufacturer A exhibits a mean attenuation of -28 dB and a steep 7.7 dB/dec roll-off up to 300 kHz. The $S_{21}$ response (Figure \ref{fig:S21_Comparison}a) also features a deep, tap-dependent notch at 150 kHz; this suggests an internal parasitic resonance that pulls the signal roughly 15 dB below the baseline. The Manufacturer B (Figure \ref{fig:S21_Comparison}b) shows a similar wavy frequency response but behaves differently at higher frequencies. Despite nearly identical mean attenuation of -26.7 dB, Manufacturer B features a shallower 5 dB/dec roll-off, and maintains tight consistency across all taps, avoiding the resonant variation seen in Manufacturer A. Both transformers are consistent up to 95 kHz, but their performance diverges significantly beyond 130 kHz. This behavior is captured by the tap-to-tap standard deviation ($\sigma$), which highlights Manufacturer A's high sensitivity to tap changes. As shown by the secondary axes in Figures \ref{fig:S21_Comparison}a \& b, the sharp resonant dips across the tap settings drive the Manufacturer A's standard deviation to a peak of 7.1 dB. In contrast, Manufacturer B remains stable, with its variation limited to just 1.3 dB. 
	
	\indent Table \ref{tab:freq_stats} also illustrates that relying on a blended, generic 25 kVA profile (the "Mfr A + B" column) limits simulation accuracy. Applying this generic profile results in a full-band standard deviation of 1.5 dB overestimates the variation of Manufacturer B ($\sigma = 0.4$ dB) while underestimating Manufacturer A's tap-driven variations. Generic models do not account for these brand-specific variations, meaning these metrics could help the researchers without access to physical measurements.  To replicate these devices in simulation, the mean attenuation at any given frequency is derived from the gain and slope entries, while the tap-dependent uncertainty is statistically generated using the spot-frequency $\sigma$ values. For example, at 150 kHz, using Manufacturer A’s metrics accounts for a $\sigma$ increase of 3.3 dB and a maximum variation of 19.6 dB. Shifting to Manufacturer B's metrics reduces that uncertainty to a $\sigma$ of 0.6 dB and a maximum variation of 3.5 dB. 
	
	\begin{table}[t]
		\centering
		\small
		\caption{Cross-Brand Statistical Indices}
		\label{tab:metrics_summary}
		\begin{tabular}{@{} l c c @{}}
			\toprule[1.5pt]
			\textbf{Metric} & \textbf{Full Band} & \textbf{Target Band} \\ 
			& (1--1601 kHz) & (45--300 kHz) \\ \midrule[1.5pt]
			Cross-Correlation Factor (CCF) & 0.38 & 0.73 \\
			Absolute Sum of Log Error (ASLE) & 3.08 dB & 1.48 dB \\ \bottomrule[1.5pt]
		\end{tabular}
	\end{table}
	
	\indent Cross-brand metrics shown in Table \ref{tab:metrics_summary} further illustrate the limitations of generic transformer models. To quantify this, the Cross-Correlation Factor (CCF) is used to assess spectral shape alignment, while the Absolute Sum of Logarithmic Error (ASLE) measures absolute magnitude differences. Across the entire 1 to 1601 kHz measurement sweep, a low CCF of 0.38 confirms that the two transformers exhibit fundamentally different resonant profiles. When isolated to the 45–300 kHz operating band, the CCF increases to 0.73 due to a shared roll-off trend. However, literature demonstrates that correlation metrics do not account for the magnitude offsets \cite{tahir2020analysis, behjat2015statistical}, meaning this shape similarity does not guarantee model accuracy. The ASLE mathematically calculates the true magnitude variation  between the two traces; this calculation shows that using a generic model in this target band still introduces a 1.48 dB baseline magnitude error. Ultimately, accurate PLC channel modeling requires manufacturer-specific data rather than standardized component profiles.
	
	\subsubsection{Auxiliary Network Components}
	Standard modeling approaches often approximate auxiliary components as ideal, transparent elements. However, the VNA characterization reveals that these components introduce non-negligible frequency-dependent behaviors that affect the channel's operational bandwidth and reflection profile.
	
	\begin{itemize}
		\item \textbf{Shunt Components:} On the LV side of the 25 kVA SWER transformer, the transmitter is deployed in a parallel (shunt) configuration alongside the LV line filter (Figure \ref{fig:TX_vs_RX_Side}a). The LV filter prevents high-frequency PLC signals from draining into the downstream customer load, forcing the injected signal energy through the transformer and onto the medium-voltage (MV) grid.
		
		\item \textbf{Series Components:} The receiver circuit located at network's feed point is illustrated in Figure \ref{fig:TX_vs_RX_Side}b). The coupling capacitor installed in series with the receiver, acts as a high-pass filter. Furthermore, the internal topology of the RX coupling circuit utilized in this receiver configuration follows the high-bandwidth design framework established in prior studies \cite{beqirllari2024high, beqirllari2025coupling}. The SWER lines's ACR is modelled as a discrete component, rather than a simple short-circuit. This captures the high-frequency insertion losses expected across the communication spectrum.
	\end{itemize}
	
	\subsection{Time-Domain Probe Signal Design}
	
\begin{figure}[t]
		\centering
		\includegraphics[width=\columnwidth]{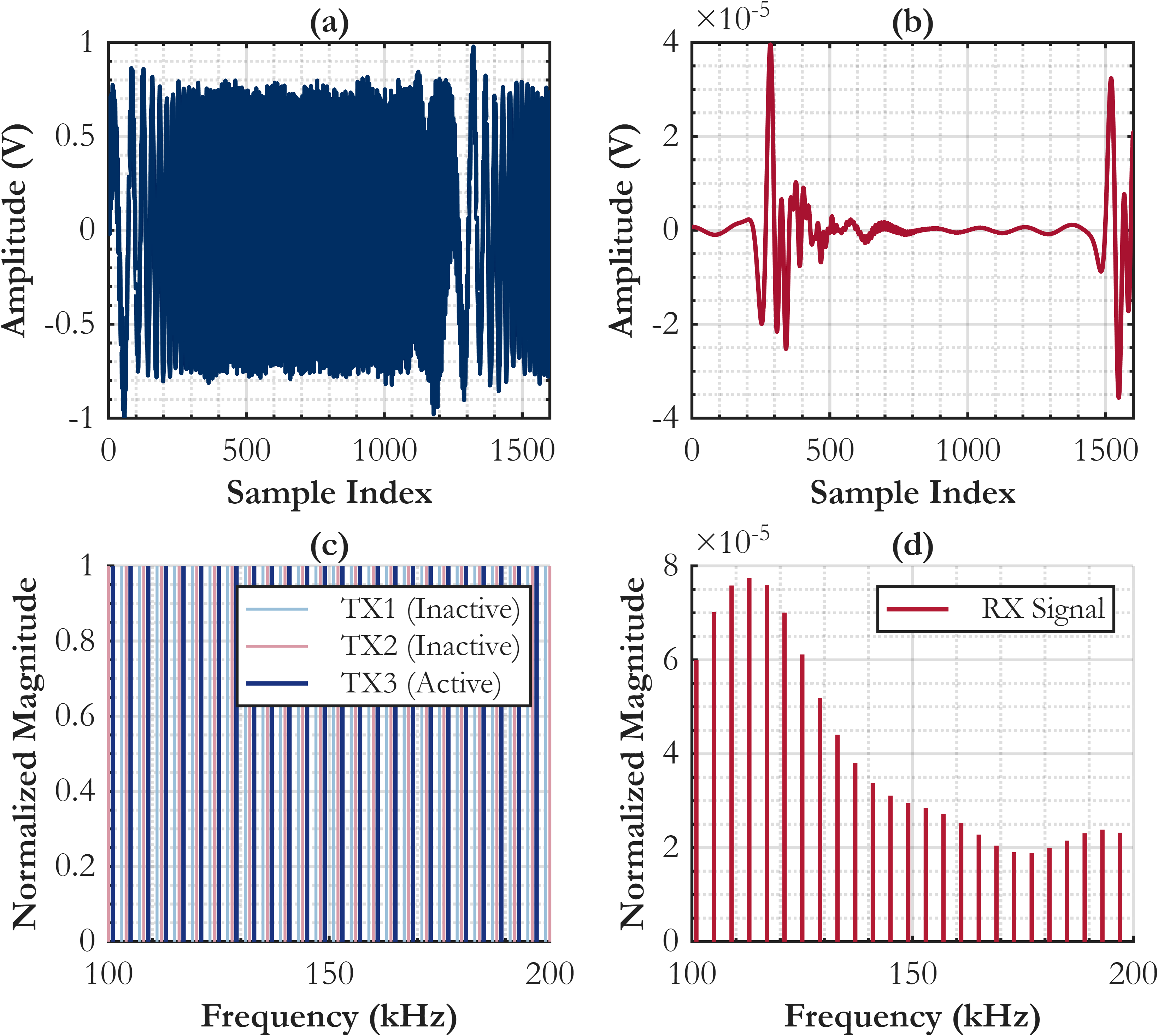}
		\caption{OFDM probing sequence characteristics for the TX3 scenario. The time-domain panels show (a) the transmitted signal ($y_{tx}$) and (b) the received signal ($y_{rx}$). The frequency-domain panels contrast (c) the transmitted interleaved Zadoff-Chu spectrum with (d) the received spectrum.}
		\label{fig:ZC}
	\end{figure}

	In the operational SWER network, the three distributed transmitters continuously transmit probe signals to monitor the channel in real time. To accurately replicate this field behavior within the Digital Twin, a signal injection architecture based on Orthogonal Frequency Division Multiplexing (OFDM) is implemented. Because all three transmitters operate simultaneously across the exact same physical infrastructure, an interleaved, comb-like subcarrier allocation strategy is employed. This orthogonal structure assigns mutually exclusive subcarrier sets to each transmitter (e.g., TX1 utilizes subcarriers $k, k+4, k+8 \dots$), as illustrated by the transmitted spectrum in Figure \ref{fig:ZC}c.
	
	To maintain a constant envelope in the time domain, a Zadoff-Chu (ZC) polyphase sequence is mapped onto these allocated subcarriers. The $k$-th element of the base ZC sequence, $Y_u(k)$, is defined in the frequency domain as:
	
	\begin{equation}
		Y_u(k) = \exp\left(-j\frac{\pi u k (k+1+2q)}{N_{ZC}}\right)
		\label{eq:ZC_equation}
	\end{equation}
	
	where $0 \le k < N_{ZC}$, $N_{ZC}$ is the sequence length (a prime number), $u$ is the root index (coprime to $N_{ZC}$), and $q$ is an arbitrary integer. By ensuring the injected signals never overlap in the frequency domain, the receiver can continuously isolate and analyze the signals arriving from each transmitter. Applying an Inverse Fast Fourier Transform (IFFT) then translates this spectrum into the time domain, producing the repeating sequence $y_{tx}$ (Figure \ref{fig:ZC}a).	
	
	\indent Because the transmitters are located at varying physical distances from the receiver (ranging from 2.1 km for TX1 to 7.5 km for TX2 and 5.7 km for TX3), the injection sequences for the more distant transmitters are proportionally amplified relative to the closest transmitter. This scaling compensates for the additional signal attenuation over the longer propagation paths. The complete SWER network model was assembled in AWR Microwave Office by cascading the transmission line segments with the empirical component data. To validate this end-to-end digital twin against the physical field setup, the synthesized transmit spectrum is multiplied by the transfer function ($H(f)$) extracted from the AWR model for each transmission path. This yields the simulated time-domain received signal ($y_{rx}$, Figure \ref{fig:ZC}b) and its corresponding received spectrum (Figure \ref{fig:ZC}d), illustrating the channel's baseline loss and frequency-selective fading.
	
	\begin{table}[t]
		\centering
		\caption{SWER Transmission Lines Parameters}
		\label{tab:SWER_Params}
		\small
		\begin{tabular}{c c} 
			\toprule[1.5pt]
			\textbf{Parameter} & \textbf{Value} \\
			\midrule[1.5pt]
			Soil Resistivity & 150 $\Omega \cdot \text{m}$ \\
			Relative Permittivity of Soil & 15\\
			Conductor Height & 6.7 m \\
			Relative Permeability of Conductor & 70\\
			\bottomrule[1.5pt]
		\end{tabular}
	\end{table}
	
	\section{Results and Analysis}
	Field measurements conducted on the SWER network using the ZC probe sequence serve as the baseline to validate the model. The co-simulation framework was initialized using RLC values derived from the equations in Section 3, with remaining environmental and material parameters detailed in  Table \ref{tab:SWER_Params}. For the sensitivity analysis, parameters were varied individually to isolate the specific impact of each variable on the channel response.
	
	\indent A preliminary comparison between the simulation and field data revealed significant discrepancies in both attenuation magnitude and spectral shape. To resolve these deviations, the analysis was structured into three stages: (1) a component-wise breakdown of channel formation, (2) a parametric sensitivity analysis of distributed and discrete variables, and (3) the final "digital twin" calibration against field data.	
	
	\subsection{Component-Wise Channel Decomposition} 
\begin{figure}[t]
		\centering
		\includegraphics[width=\columnwidth]{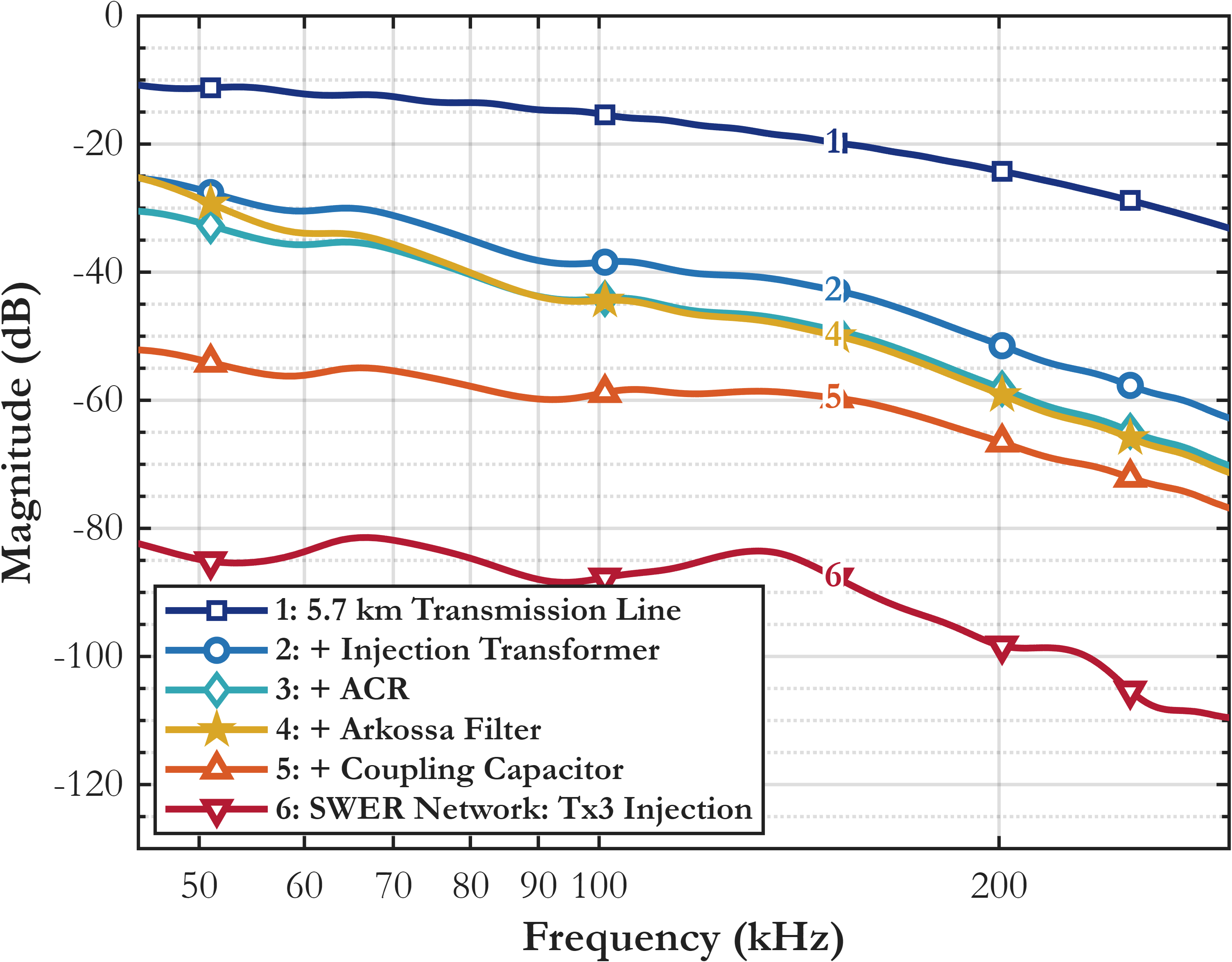}
		\caption{Step-wise decomposition of the received voltage spectrum, illustrating the cumulative gain of the OFDM probe signal caused by the transmission line (TX3, 5.7km), discrete components, and full network topology.}
		\label{fig:cumulative_loss}
	\end{figure}

To rigorously quantify the contribution of individual network elements to the total link budget, a constructive simulation experiment was performed. A simplified 5.7 km linear baseline representing the TX3 to RX path (excluding network spurs) was modeled, into which discrete components were progressively inserted. Figure \ref{fig:cumulative_loss} illustrates the cumulative impact of these elements on the received voltage spectrum. 
	
	\indent The baseline 5 km SCGZ conductor (curve 1) exhibits a linear, predictable voltage decay governed by the skin effect and earth return path, with the received signal magnitude smoothly dropping from approximately $-10$ dB to $-35$ dB across the band. However, the insertion of the primary injection transformer (curve 2) introduces an immediate, massive signal drop of 15 to 30 dB. This single discrete component degrades the received voltage significantly more than the entire 5.7 km transmission line.
	
	\indent The subsequent addition of the series and shunt isolation components further reshapes the received voltage spectrum. The ACR (curve 3) introduces a consistent incremental drop, lowering the voltage magnitude by approximately 4 to 6 dB across the spectrum. Crucially, the addition of the line filter (curve 4) introduces almost negligible penalty to the communication channel. This filter is physically deployed to block high-frequency PLC signals from sinking into the customer load while still allowing the standard 50 Hz mains power to pass. Across the higher frequencies, its response virtually overlaps the cumulative ACR curve. In contrast, the series coupling capacitor (curve 5) demonstrates a distinct high-pass filter response. Physically designed to block the high-voltage mains power while safely extracting the communication signals, it acts as a severe bottleneck at lower frequencies, suppressing the signal magnitude by over 20 dB near 50 kHz. However, as the frequency increases, its series impedance drops accordingly, narrowing the attenuation gap to approximately 5 dB at the upper end of the communication spectrum.
	
	\indent The full SWER network simulation for the TX3 injection scenario (curve 6) reveals the critical impact of topological interaction. While the unbranched linear path (curve 5) results in a received voltage between $-52$ and $-75$ dB, the full network topology results in a signal attenuation of $-82$ to $-110$ dB range. This significant suppression ($>25$ dB) is driven by multipath energy splitting and the aggregate impedance of the multi-branched network. The complex network topology presents a vastly different driving impedance compared to the unbranched direct path, establishing the ultimate, highly attenuated voltage floor for the communication channel.
	
	\indent Finally, the impact of downstream MV-to-LV transformer tap settings was evaluated. While the primary injection transformer directly controls signal throughput, downstream transformers act as reflective boundaries at the end of the line. As shown in Figure \ref{fig:S22_Comparison}, their MV-side reflection ($S_{22}$) remains almost constant across all tap positions. Because this reflection does not change, adjusting the taps has almost no effect on the network's standing waves. Testing the full range of downstream taps altered the total attenuation by less than 2 dB and did not shift the resonant notches or change the overall frequency response. Since their impact is minimal, downstream tap variations are excluded from the main analysis to keep the results clear.
	
	\subsection{Parametric Investigation of the SWER Channel Model}
	The initial co-simulation model of the complete network was configured using standard baseline parameters: the injection transformer (Manufacturer A) was set to Tap 5, the conductor relative permeability to $\mu_r = 70$, and the environmental variables matched those in Table \ref{tab:SWER_Params}. However, early comparisons with the field measurements showed noticeable differences in both the overall attenuation and the shape of the frequency response. To identify the origin of this discrepancy, a parametric sensitivity analysis was performed. This isolated the individual effects of soil resistivity, transformer tap settings, conductor height, and magnetic permeability to determine exactly which factors dictate the channel's spectral signature.
	
	\begin{figure}[t]
		\centering
		\includegraphics[width=\columnwidth]{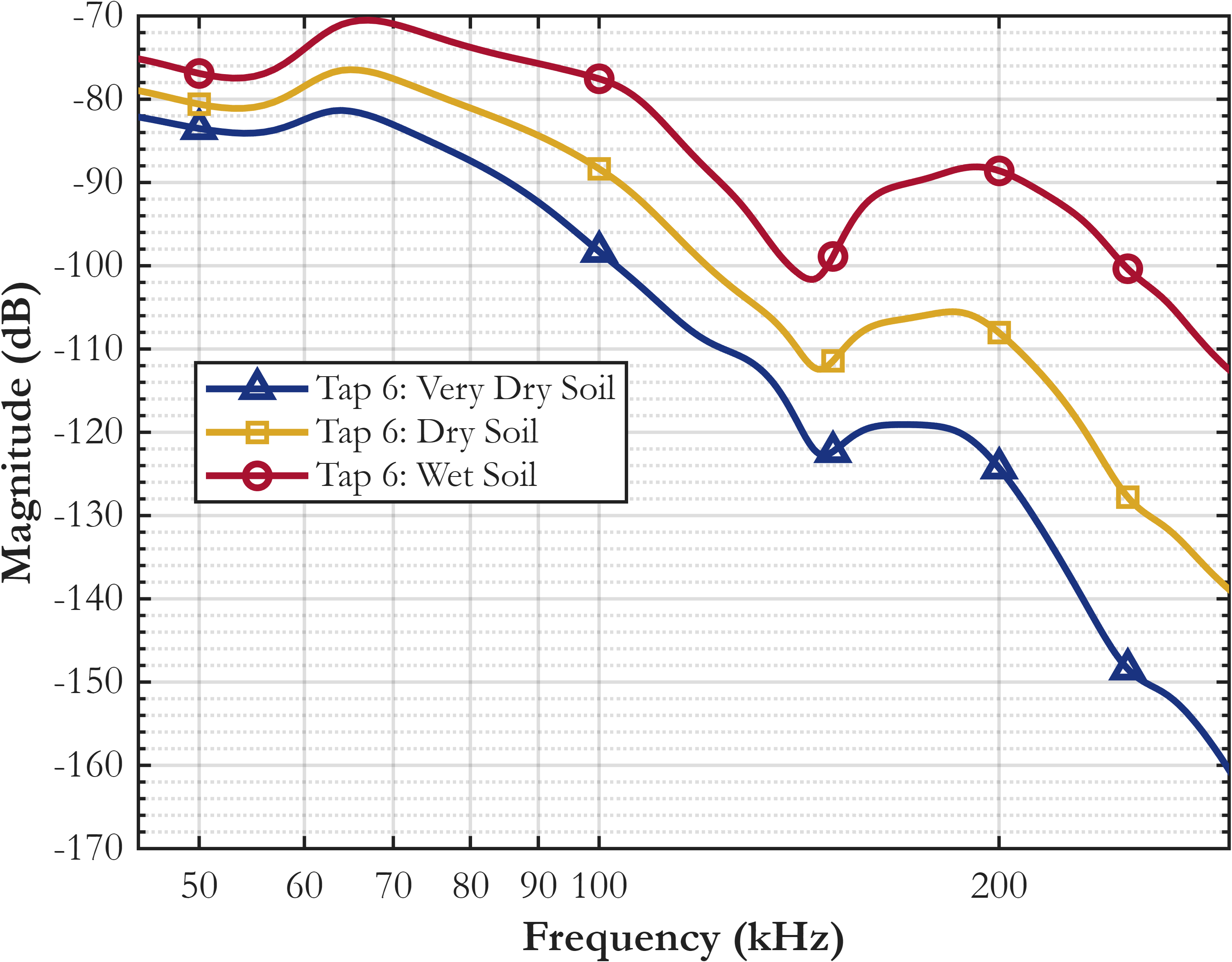}
		\caption{Sensitivity analysis of the TX2 channel response illustrating the effect of varying soil resistivity.}
		\label{fig:soil_sensitivity}
	\end{figure}
	
	\begin{figure*}[t]
		\centering
		\begin{subfigure}[b]{0.49\textwidth}
			\centering
			\includegraphics[width=\textwidth]{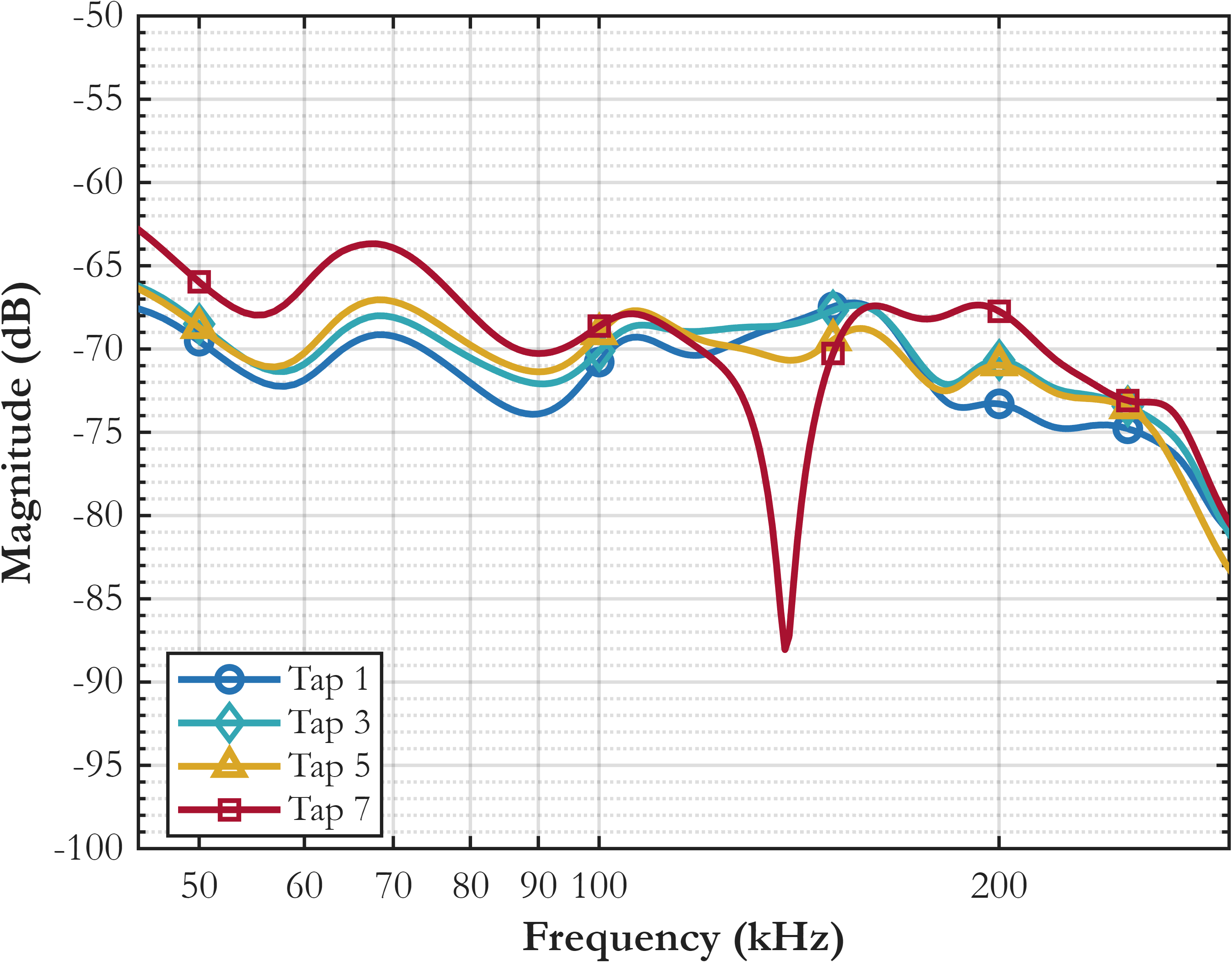} 
			\caption{TX1--2.1 km from RX}
			\label{fig:tap_tx1}
		\end{subfigure}
		\hfill
		\begin{subfigure}[b]{0.49\textwidth}
			\centering
			\includegraphics[width=\textwidth]{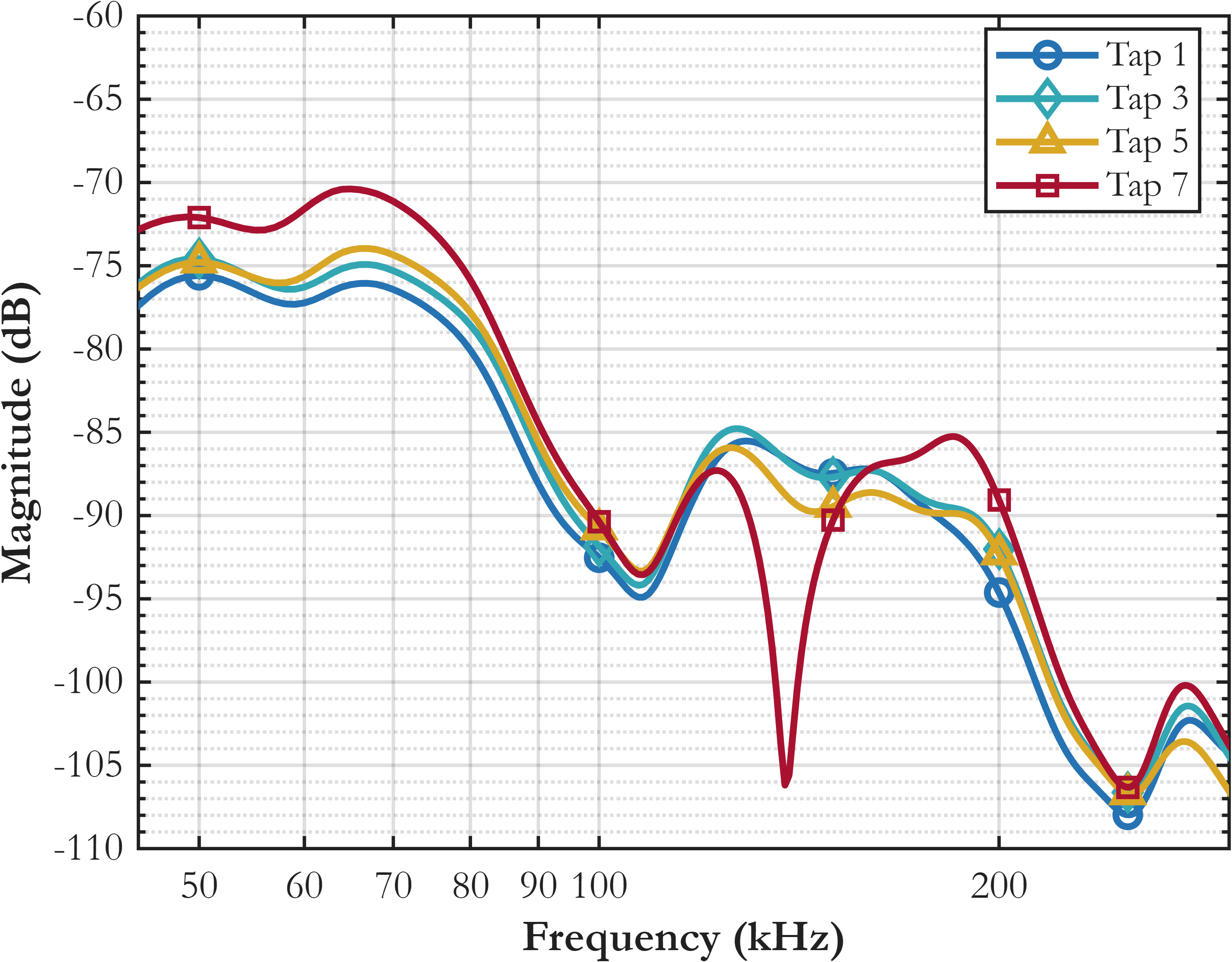} 
			\caption{TX2--7.5 km from RX}
			\label{fig:tap_tx2}
		\end{subfigure}
		
		\caption{Impact of transformer tap settings on the received voltage magnitude. (a) Received signal from TX1 (b) Received signal from TX2.}
		\label{fig:tap_sensitivity}
	\end{figure*}
	
	\subsubsection{Sensitivity to Earth Return Path} Since SWER systems rely on the earth as the return path, the communication channel's performance is heavily influenced by soil properties. To determine exactly how much the soil type matters, the TX2 injection path was simulated across a spectrum of standardized soil profiles. These profiles are derived from the ground classifications established in ITU-R Recommendation  \cite{itu_p527_6}. The analysis compares a realistic baseline of Medium dry ground ($\rho = 1000\,\Omega\cdot\text{m}, \epsilon_r = 15$) against two boundary extremes: Very dry ground ($\rho = 10000\,\Omega\cdot\text{m}, \epsilon_r = 3$) and Wet ground ($\rho = 100\,\Omega\cdot\text{m}, \epsilon_r = 30$).
	
	\indent As Figure \ref{fig:soil_sensitivity} shows, the soil moisture has a massive impact on the overall signal magnitude. This effect is especially pronounced at higher frequencies; while the difference in received voltage is relatively small at 50 kHz, the gap widens significantly as frequency increases, reaching a difference of approximately 30 dB near 200 kHz. Physically, this is tied to the frequency-dependent earth return impedance ($Z_e$), modeled here using Sunde's logarithmic approximation. At higher frequencies, the skin effect forces the return current to flow closer to the surface. Because the current is confined to a shallower path, it becomes far more sensitive to surface resistivity. Lowering the soil resistivity shrinks the resistive component of $Z_e$ and pulls this return current path even closer to the surface. This tightens the loop inductance of the SWER circuit and drastically reduces propagation losses, confirming that the earth return determines the overall attenuation slope of SWER channel model.
	
	\indent Importantly, the actual shape of the spectrum remains remarkably stable across these different soil types, specifically regarding where the resonant peaks and dips occur. Even though the overall received voltage shifts vertically by dozens of decibels, these spectral dips only move by about 2 to 5 kHz. This proves that while soil conditions set the baseline attenuation floor, they do not significantly alter the overall channel response.
	
	\subsubsection{Sensitivity to Transformer Tap Selection}
	As illustrated in Figure \ref{fig:S21_Comparison}, varying the transformer tap setting changes its internal leakage inductance and inter-winding capacitance. When applied to the full network simulation (Figure \ref{fig:tap_sensitivity}), this alters the overall channel response, physically displacing the spectral dips for both the TX1 and TX2 scenarios. The dominant mechanism is the forward transfer function ($S_{21}$), changing the signal injected into the network. This is evident from the relative difference between the taps that mirrors those of Figure \ref{fig:S21_Comparison}a. 
	
	\indent Comparing the two injection nodes also highlights the significant impact of physical distance. For the shorter TX1 path (2.1 km), the channel response stays relatively steady even at higher frequencies, because the signal reaches the receiver before high-frequency losses can significantly compound. In contrast, the TX2 path (7.5 km) shows a severe drop-off in the high-frequency spectrum above 150 kHz. This behavior is primarily driven by the physics of the transmission line. Skin effect and earth-return losses naturally increase with frequency, resulting in higher 'dB per meter' attenuation in the upper band. Additionally, the longer the signal travels through the network, it encounters more branches and boundaries that create additional reflections, leading to deeper frequency-selective fading.
	
	\subsubsection{Sensitivity to Conductor Height}
	Another practical structural factor in SWER systems is the height of the overhead conductor. Variations in pole design and seasonal line sag change the physical distance between the wire and the earth. To evaluate how this impacts the communication channel, we tested the TX2 injection path using the baseline conductor height ($h$), a severe sag condition ($0.75h$), and an elevated profile ($1.8h$).
	
	\begin{figure}[t]
		\centering
		\includegraphics[width=\columnwidth]{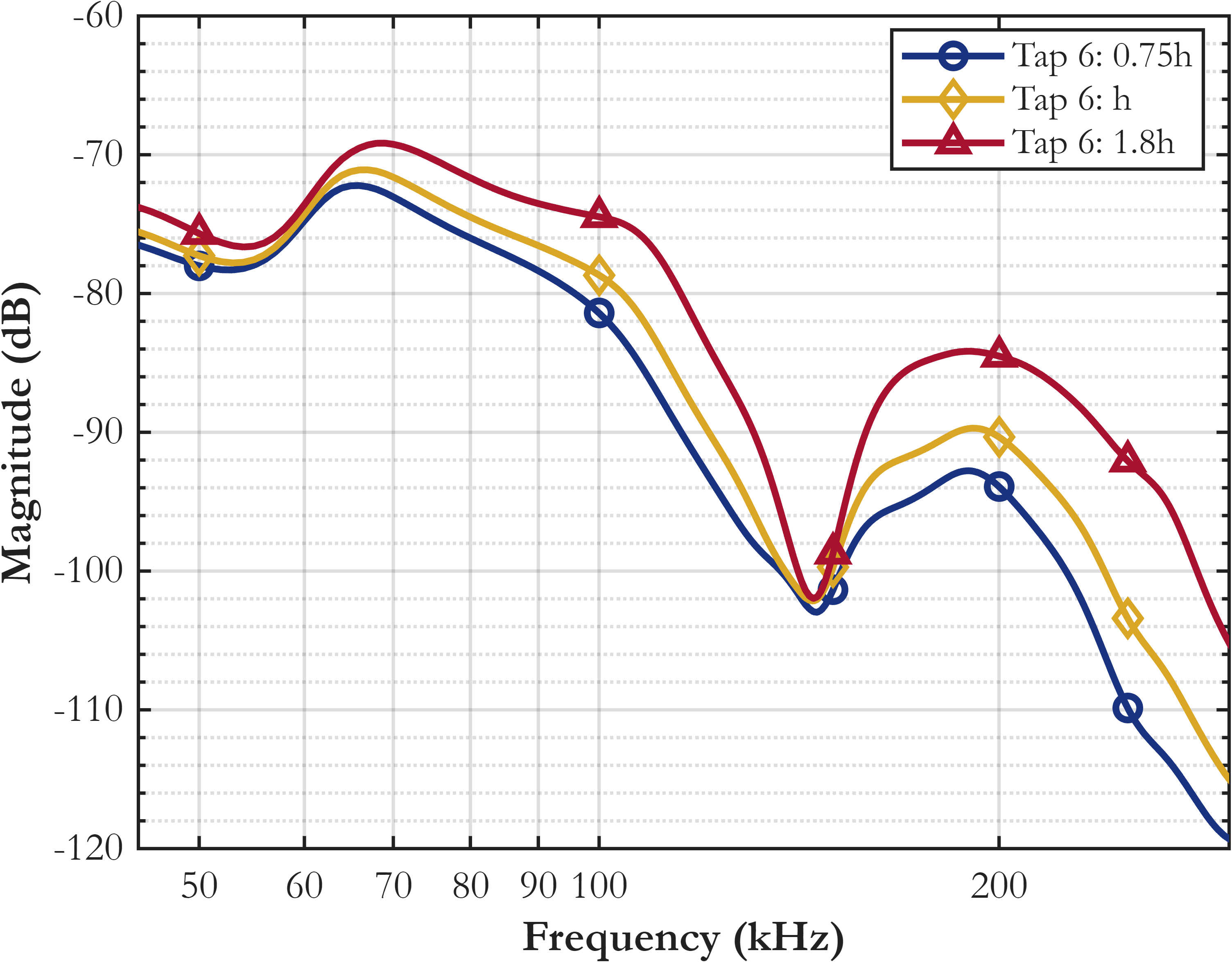}
		\caption{Sensitivity analysis of the TX2 channel response illustrating the effect of varying overhead conductor height.}
		\label{fig:height_sensitivity}
	\end{figure}
	
	\indent As shown in Figure \ref{fig:height_sensitivity}, the conductor's height has a direct and significant effect on signal attenuation. The elevated $1.8h$ line preserves the most signal, while the sagging $0.75h$ line suffers the greatest losses. Physically, this is driven by the shunt capacitance between the power line and the ground. Raising the conductor increases the air gap, which decreases the capacitance and increases the inductance. Consequently, the characteristic impedance ($Z_0 \approx \sqrt{L/C}$) of the elevated line increases. As defined in Equation 11, the attenuation constant is inversely proportional to the magnitude of the characteristic impedance. Therefore, the higher $Z_0$ of the elevated line inherently results in lower signal attenuation. Furthermore, this relationship explains the widening performance gap at higher frequencies. Due to severe skin effect and earth-return losses, the total resistance ($R_{total}$) rises sharply in the upper spectrum. The lower $Z_0$ of the sagging line amplifies the impact of this high-frequency resistance, causing the magnitude gap between the $0.75h$ and $1.8h$ profiles to widen to more than 10 dB above 200 kHz.
	
	\indent Importantly, changing the conductor height does not alter the underlying shape of the spectrum. The spectral dips remain locked at the same frequencies across all three height scenarios. This reinforces the earlier observation: physical variations like gradual line sag will impact the baseline attenuation floor by altering the line's inherent loss profile, but they do not shift the network's overall resonant conditions. 
	
	\subsubsection{Sensitivity to Conductor Permeability} 
	SWER networks frequently utilize steel-cored conductors to achieve long span lengths; therefore, the magnetic properties of the transmission line play a critical role in high-frequency signal propagation. To evaluate this, we swept the relative magnetic permeability ($\mu_r$) of the conductor from 70 to 200 on the TX2 path.
	
\begin{figure}[t]
		\centering
		\includegraphics[width=\columnwidth]{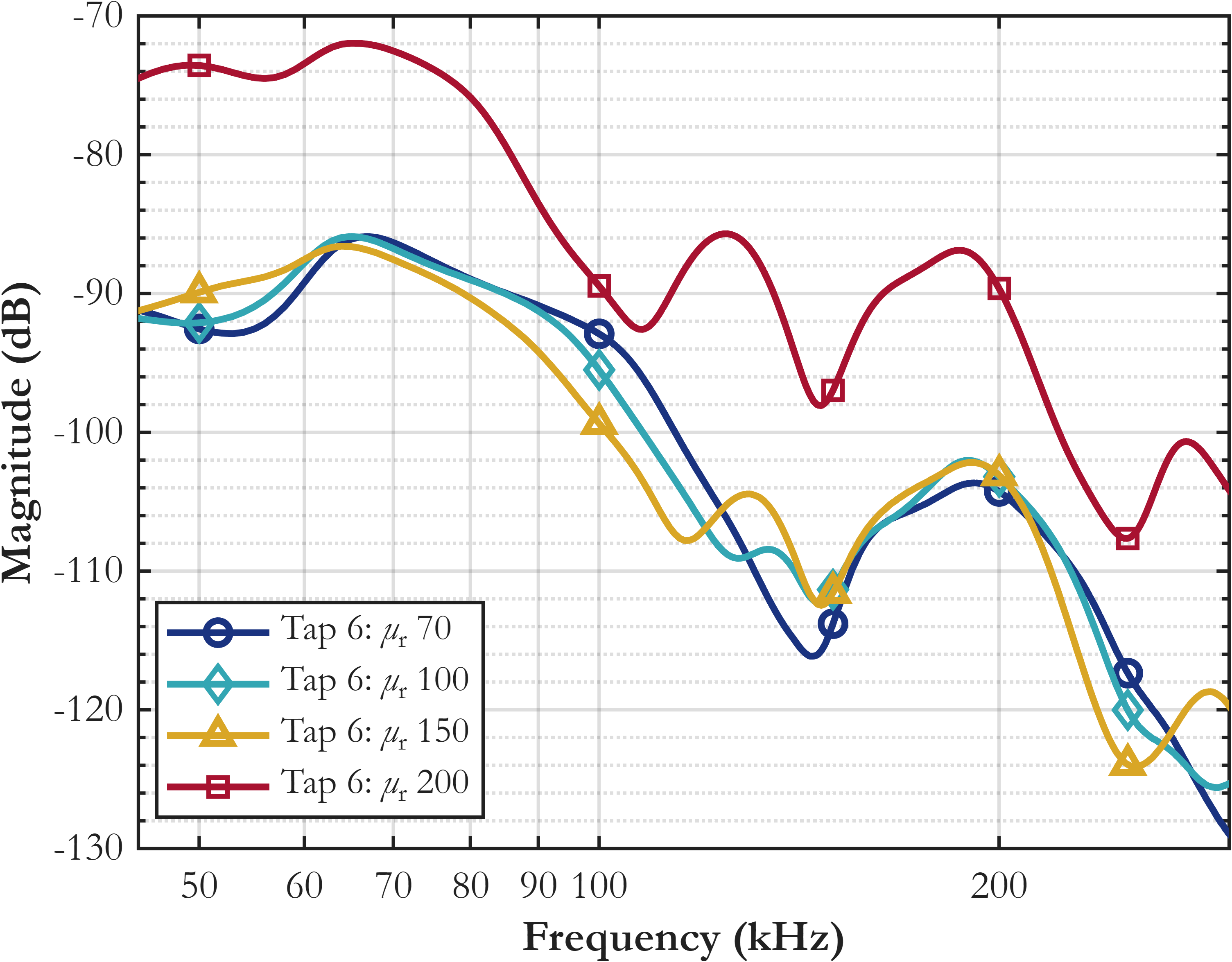}
		\caption{Sensitivity analysis of the TX2 channel response to relative permeability.}
		\label{fig:permeability_sensitivity}
	\end{figure}

	\indent As shown in Figure \ref{fig:permeability_sensitivity}, varying the magnetic permeability produces a drastically different effect compared to soil moisture or conductor sag. Instead of just shifting the overall signal magnitude up or down, changing $\mu_r$ fundamentally alters the actual shape of the frequency spectrum. While the channel responses for $\mu_r$ 70, 100, and 150 remain somewhat clustered with minor variations in attenuation, stepping up to $\mu_r$ 200 reshapes the spectrum. The overall magnitude increases, and the spectral dips are also displaced.
	
\begin{figure}[t]
		\centering
		\includegraphics[width=\columnwidth]{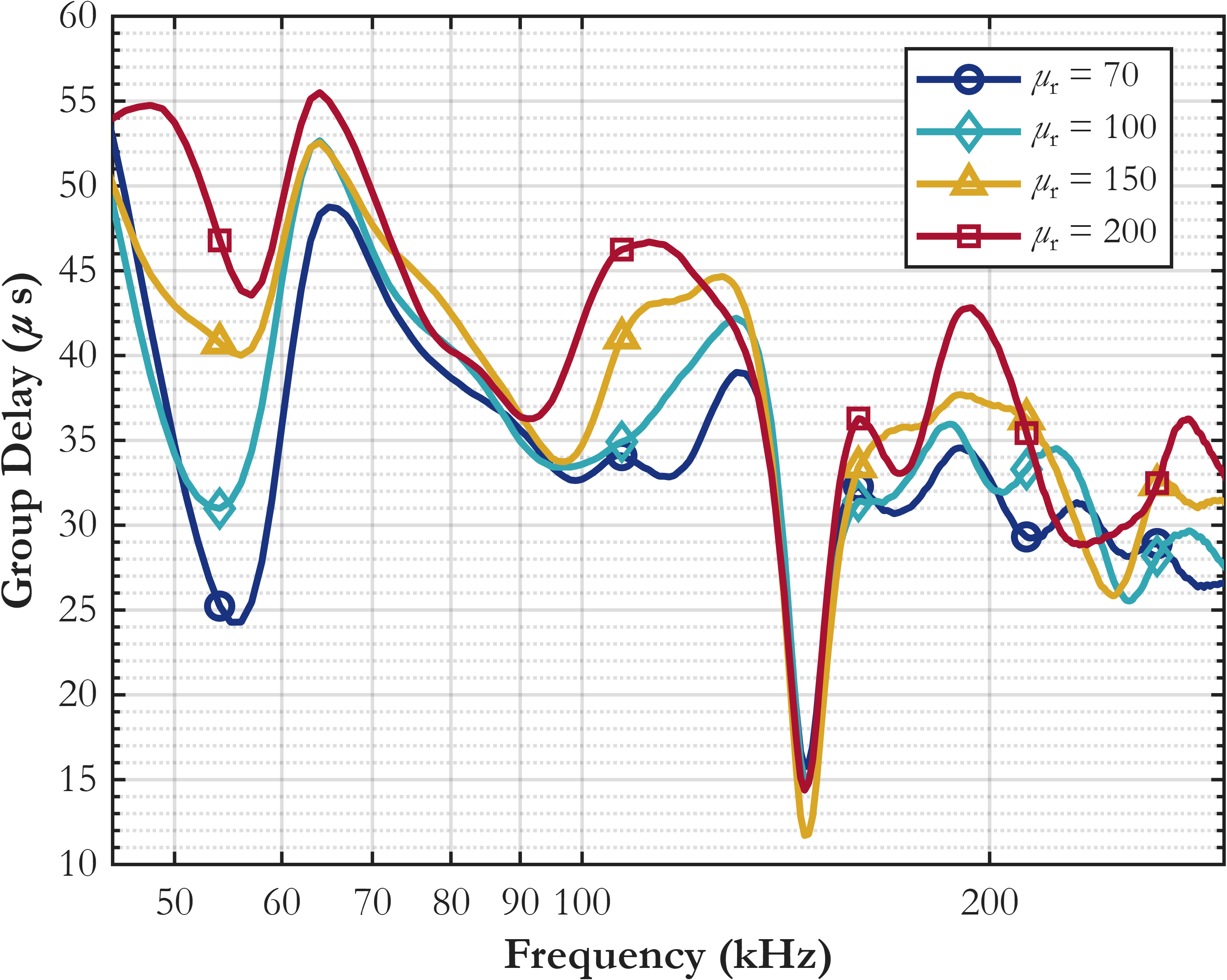} 
		\caption{Group delay profile of the TX2 channel under varying conductor permeability, demonstrating increased propagation time at higher $\mu_r$ values.}
		\label{fig:group_delay_sensitivity}
	\end{figure}

	\indent The mechanism driving this shape change is the velocity of signal propagation and the line impedance. The relative permeability of the steel core directly dictates its internal inductance ($L_{int}$); as $\mu_r$ increases, the total series inductance ($L$) of the line also increases, as does $Z_0$, reducing attenuation. Because transmission line wave velocity ($v$) is inversely proportional to the square root of inductance ($v \approx 1/\sqrt{LC}$), the high permeability of the $\mu_r = 200$ conductor physically slows down the high-frequency carrier wave, changing the multipath interference pattern. This reduction in velocity is directly captured by the group delay ($\tau_g$) as illustrated in Figure \ref{fig:group_delay_sensitivity}, which represents the time required for a signal envelope to traverse the network. Mathematically, group delay is the negative derivative of the phase response with respect to angular frequency ($\tau_g = -d\phi/d\omega$). As the wave velocity drops due to increased inductance, the total propagation time across the 7.5 km network necessarily increases. 
	
	\indent Consequently, because the wave travels more slowly through the network, its spatial wavelength ($\lambda = v/f$) is compressed. Because these wavelengths are physically shorter, a greater number of standing wave cycles can fit into a given length of the transmission line. This compression forces the standing wave nulls to occur more frequently across the spectrum, essentially squeezing the resonant pattern together. This is clearly visible in Figure \ref{fig:permeability_sensitivity}. Rather than just shifting uniformly to the left, the $\mu_r = 200$ profile packs a higher density of spectral notches into the communication band than the $\mu_r = 70$ baseline. Interestingly, both Figure \ref{fig:permeability_sensitivity} and Figure \ref{fig:group_delay_sensitivity} show a sharp, deep resonance notch near 148 kHz that remains completely stationary across all permeability sweeps. This provides visual validation for the earlier component-wise decomposition: this specific attenuation null is a fixed hardware characteristic of the injection transformer, whose internal resonance remains strictly independent of the distributed transmission line parameters. Ultimately, this proves that the magnetic properties of the conductor dictate the spatial density of the standing waves across the channel.
	
	\subsection{Field Validation and Digital Twin Calibration}
	Following the sensitivity analysis, the digital twin was calibrated to align with field measurements by tuning the two parameters demonstrated to have the greatest impact on signal propagation: the injection transformer tap setting and the conductor's relative magnetic permeability ($\mu_r$). As established in the network heterogeneity analysis (Section 3), the transformers from different manufacturers exhibit distinct spectral signatures. For this validation against the physical SWER network, the transformer from manufacturer A was selected as the baseline injection element. 
	
	\begin{figure}[t]
		\centering
		\includegraphics[width=\columnwidth]{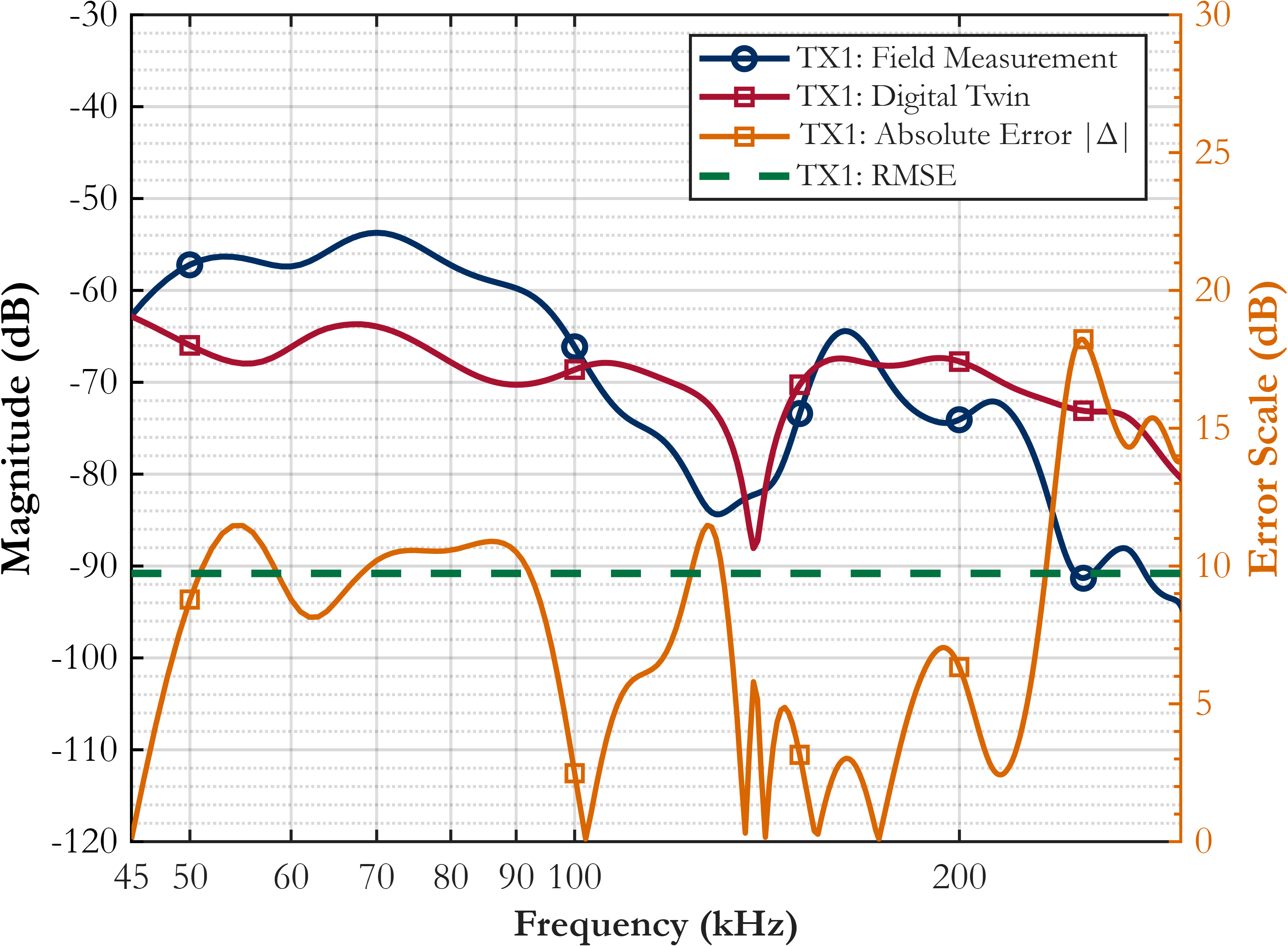}
		\caption{TX1 Validation: Field Measurement vs. Simulation (Tap 7).}
		\label{fig:validation_tx1}
	\end{figure}
	
	\indent The relative magnetic permeability of galvanized steel cores covers a diverse range. As established in transmission line literature, the effective $\mu_r$ of a steel core depends on the alloy composition, manufacturing process, prevailing magnetic-field strength, the high-frequency skin effect, ambient temperature, and the physical tensile stress of the line \cite{morgan2013current}. Because the exact in-situ mechanical tension, magnetic history, and high-frequency alloy response of the physical network cannot be perfectly known, $\mu_r$ was treated as a necessary calibration parameter. Sweeping this parameter revealed that $\mu_r = 200$ provided the optimum alignment and spectral matching across all three field sites. 
	
	\begin{figure}[t] 
		\centering 	
		\includegraphics[width=\columnwidth]{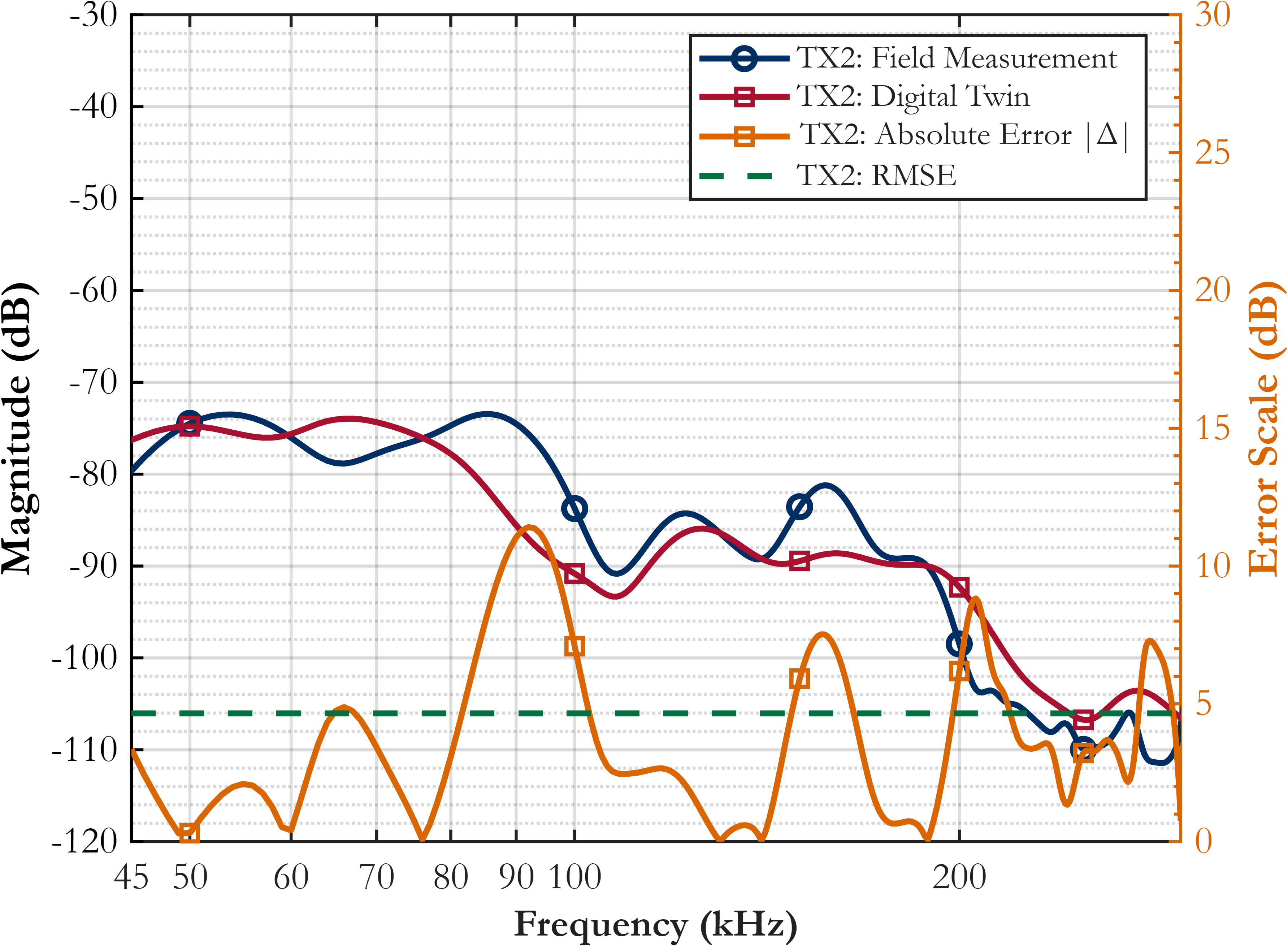} 
		\caption{TX2 Validation: Field Measurement vs. Simulation (Tap 5).} 
		\label{fig:validation_tx2} 
	\end{figure}
	
	\indent \indent With regards to the tap settings, utilities routinely adjust these at individual nodes to compensate for localized voltage drops across long radial distances. Because the exact tap positions at the time of the field measurements were not explicitly logged, they were treated as bounded calibration parameters, restricted to the physical Tap 1 through Tap 7 range of the Manufacturer A transformer. The optimum spectral matches were achieved using Tap 7 for TX1, Tap 5 for TX2, and Tap 3 for TX3. This variance realistically reflects the localized voltage regulation required across different physical points of the network.
	
\begin{figure}[t] 
		\centering 	
		\includegraphics[width=\columnwidth]{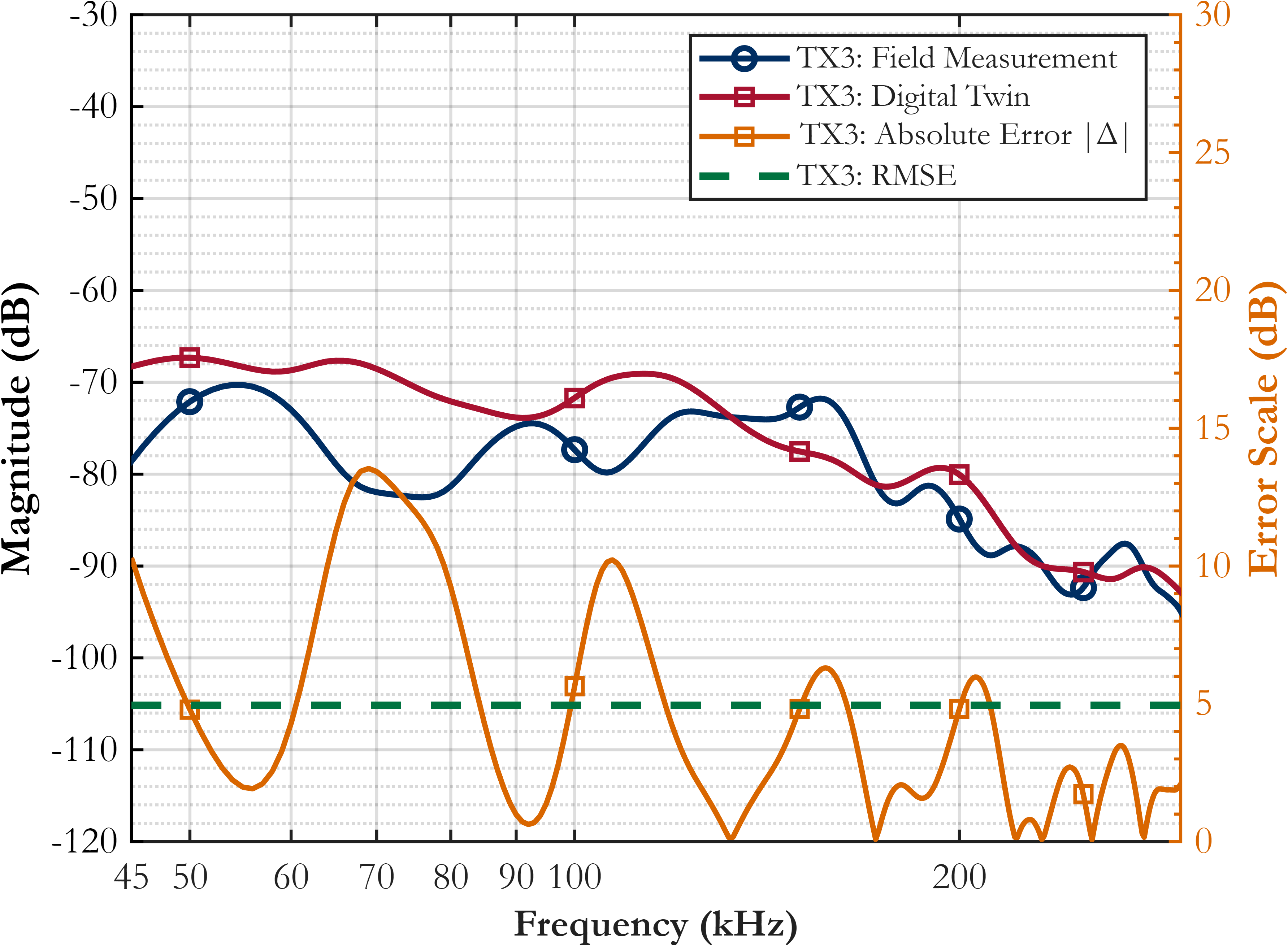} 
		\caption{TX3 Validation: Field Measurement vs. Simulation (Tap 3).} 
		\label{fig:validation_tx3} 
	\end{figure}

	\indent As shown in the validation plots, the calibrated digital twin successfully replicates the complex field measurements for all three transmitters. For the TX1 path (Figure \ref{fig:validation_tx1}), the simulation successfully captures the dominant spectral dip near 140 kHz, alongside the high-frequency roll-off trajectory, yielding a Root Mean Square Error (RMSE) of 9.73 dB (Table \ref{tab:validation_metrics}). The TX2 and TX3 validations further highlight the model's reliability. For the TX2 path (Figure \ref{fig:validation_tx2}), the simulation successfully recreates the rapid changes in peaks and dips between 100 kHz and 300 kHz, resulting in an RMSE of 4.65 dB. The TX3 response (Figure \ref{fig:validation_tx3}) also provides a close match; the digital twin underestimates the path loss by about 4 dB for frequencies below 120 kHz, before closely following the measured results at higher frequencies, achieving an overall RMSE of 4.95 dB.
	
	\begin{table}[t]
		\centering
		\caption{Digital Twin Validation Metrics}
		\label{tab:validation_metrics}
		\footnotesize
		\setlength{\tabcolsep}{4.5pt} 
		\begin{tabular}{@{} l c c c @{}} 
			\toprule[1.5pt]
			\textbf{Transmitter} & \textbf{RMS Diff. (dB)} & \textbf{Max Error (dB)} \\
			\midrule[1.5pt]
			TX1 & 9.73 & 18.23 \\
			TX2 & 4.65 & 11.41 \\
			TX3 & 4.95 & 13.54 \\
			\bottomrule[1.5pt]
		\end{tabular}
	\end{table}
	
	\indent These deviations are primarily attributable to two physical realities. First, transient environmental variables such as daily fluctuations in localized soil moisture, ambient temperature variations affecting conductor sag, and unmapped vegetative encroachment can subtly shift the baseline attenuation. Second, the physical network possesses an inherent degree of equipment heterogeneity. Mapping the exact parameters of every physical network component is neither practical nor computationally feasible. Crucially, however, the digital twin captures the underlying shape of the frequency spectrum at all three sites. By accurately replicating the frequency-selective fading and the overall resonant conditions across multiple network branches, the model provides a digital replica of the physical SWER architecture.
	
	\section{Conclusion}
	This paper presented the first segment-by-segment digital twin of a SWER network, directly validated against  field measurements. The step-wise channel decomposition yielded several critical insights for PLC deployment. A major quantitative finding is that the injection transformer is the primary source of channel attenuation. This indicates that optimizing transformer coupling offers the greatest return on investment for improving overall PLC performance. Furthermore, the analysis proved that the full branched network topology suppresses the signal floor significantly more than a simple linear path. This confirms that existing point-to-point models are inadequate for designing SWER PLC links, making full network analysis an absolute requirement. The study also highlighted the critical role of network heterogeneity; transformers of identical ratings from different manufacturers produced fundamentally different spectral signatures, confirming that generic component models are insufficient for accurate simulation.
	
	\indent Beyond component impacts, the parametric analysis established a clear physical distinction between variables that shift signal strength and those that distort spectral shape. Environmental and structural factors, such as soil moisture and conductor sag, shift the overall signal magnitude vertically but leave the underlying spectral shape intact. In contrast, equipment configurations and conductor's material properties fundamentally alter the resonant conditions of the channel. Specifically, transformer tap settings modify the complex source impedance at the injection boundaries, while the magnetic permeability of the steel core dictates the internal inductance and, consequently, the signal propagation velocity. Together, these parameters physically displace the multipath peaks and dips to entirely new frequencies.  
	
	\indent Ultimately, validating the digital twin against three physical transmit sites confirms it is an accurate replica of the operational SWER network. By successfully tracking the complex frequency fading of the live grid, yielding RMSE between 4.65 dB and 9.73 dB across the three transmit paths, the model proves to be a dependable representation of the operational SWER network. Moving forward, this digital twin can be used to simulate fault conditions, evaluate new communication protocols, and optimize smart grid deployments without the expense or risk of physical field trials.
	
	\section*{Acknowledgments}
	This work was supported by the Australian Research Council (ARC) Industry Fellowship (IE230100635). This research was conducted by the VU Power Sensing Group as part of the ongoing SWER Broken Conductor project, an industry-supported initiative aimed at enhancing rural distribution network safety and reliability.
	
	\bibliographystyle{elsarticle-num}
	\bibliography{library}
	
\end{document}